\documentclass[twocolumn]{aastex631}

\usepackage{amsmath}
\usepackage{multirow}
\usepackage{xcolor} 
\usepackage{hyperref}

\begin{document}

\title{Population demographics of post-interaction WDMS binaries: From common envelope evolution to stable mass transfer}

\author[0000-0001-6970-1014]{Natsuko Yamaguchi}\footnote{corresponding author (nyamaguc@caltech.edu)}
\affiliation{Department of Astronomy, California Institute of Technology, 1200 E. California Blvd, Pasadena, CA, 91125, USA}

\author[0000-0002-6871-1752]{Kareem El-Badry}
\affiliation{Department of Astronomy, California Institute of Technology, 1200 E. California Blvd, Pasadena, CA, 91125, USA}

\author[0000-0003-1247-9349]{Cheyanne Shariat}
\affiliation{Department of Astronomy, California Institute of Technology, 1200 E. California Blvd, Pasadena, CA, 91125, USA}

\begin{abstract}
Close white dwarf (WD) + main-sequence (MS) binaries are end products of mass transfer (MT) that occurred prior to WD formation, making their population demographics a powerful probe of binary evolution. Several recent works have constructed samples of WD+MS binaries with well-understood selection functions using data from wide-field surveys. These include (a) AU-scale astrometric binaries from Gaia that can be shown to contain a WD on dynamical grounds, (b) AU-scale astrometric binaries in which a hot WD is detected through a GALEX UV excess, and (c) close binaries discovered through eclipses. Together, these samples probe outcomes of both stable MT and common-envelope evolution, and interactions on both the red giant branch (RGB) and asymptotic giant branch (AGB). We forward model the three observed samples simultaneously. This approach produces robust constraints on uncertain binary evolution parameters because binaries removed from one population are predicted to appear in another. Our modeling includes a realistic initial binary population and treatments of the selection effects affecting all samples. We confirm that MT from AGB donors requires a critical accretor-to-donor mass ratio of $\sim0.4$ as found in previous work, and find that this is more stable than MT from RGB donors, for which we constrain a critical ratio $\gtrsim0.65$. A common envelope efficiency of $\alpha\lambda\sim0.3$ matches the relative numbers of close and wide systems and the period distribution of close systems. Most stable MT products in the sample, including those with RGB donors, retain nonzero eccentricities ($\simeq0.1$). The model does not fully reproduce the mass distribution of main-sequence stars in post-common envelope binaries, which shows a cliff below the fully convective limit, possibly pointing to missing physics that may warrant future work.
\end{abstract}

\keywords{Binary stars (154) --- White dwarf stars (1799) --- Astrometry (80)}

\section{Introduction} \label{sec:intro}

\defcitealias{Shariat2026PASP}{SE26}

Binary mass transfer (MT) plays a critical role in the formation of a wide range of important and exotic astrophysical phenomena, including millisecond pulsars \citep[e.g.][]{Radhakrishnan1982CSci, Alpar1982Natur, Bhattacharya1991PhR}, Type Ia supernovae \citep[e.g.][]{Whelan1973ApJ, Nomoto1982ApJ, Iben1984ApJS, Webbink1984ApJ}, X-ray binaries \citep[e.g.][]{Gursky1971ApJL, Rappaport1982ApJ}, and gravitational wave sources \citep[e.g.][]{Peters1963PhRv, Hulse1975ApJL, Abbott2016PhRvL}. 

One approach to constraining MT physics is to study the end products of MT and work backwards. Many binaries hosting white dwarfs (WD) in tight ($a\lesssim10\,$AU) orbits experienced MT when the WD progenitor was on the red giant branch (RGB) or asymptotic giant branch (AGB) and overfilled its Roche lobe. Depending on the initial separation and mass ratio, this MT may have been stable or unstable. Stable MT generally leads to wide orbits, while unstable MT leads to common envelope evolution (CEE) and significant orbital shrinkage \citep[e.g.][]{Willems2004A&A}. In this work, we study binaries containing a WD and a luminous, nondegenerate companion -- usually, a main-sequence (MS) star. We refer to such systems as WDMS binaries, though we note that the companion can also be evolved or temporarily inflated (e.g. \citealt{Bhattacharjee26}).

While reconstructing the detailed evolutionary history of any one binary is challenging due to model degeneracies, the parameter distributions (i.e., orbital periods, component masses, and eccentricities) of a large sample of post-MT WDMS binaries can provide meaningful population-level constraints on the MT process. Leveraging population demographics to constrain MT requires homogeneously selected samples with well-understood selection functions. Ideally, observed samples should also span a wide range of orbital parameters and probe diverse MT histories. Much of the previous work on post-interaction WDMS binaries has analyzed heterogeneous samples with selection functions that were difficult to characterize (e.g. \citealt{Green1986ApJS, Schreiber2003A&A, Liebert2005ApJS, Rebassa-Mansergas2010MNRAS, Rebassa-Mansergas2012MNRAS}). This has made it challenging to infer intrinsic population demographics of WDMS binaries, and particularly to compare post-stable MT and post-CEE systems on equal footing.

Two datasets have recently enabled a systematic study of the WDMS binary population on an unprecedented scale. First, astrometry from the 3rd data release of the Gaia mission \citep{GaiaCollaboration2023A&A, GaiaCollaboration2023A&Ab} provided orbital solutions for more than 160,000 AU-scale binaries ($P_{\rm orb}\sim 100-1000\,$d), including thousands of post-interaction systems containing WDs \citep{Shahaf2024MNRAS}. Second, light curves from the Zwicky Transient Facility (ZTF; \citealt{Bellm19, Graham19, Masci19}) now provide $\sim$ 1000 photometric measurements for more than a billion stars. These light curves can reveal short-period ($P_{\rm orb}\lesssim 2\,$d) WDMS binaries via eclipses.

%But with the third data release of the Gaia mission (Gaia DR3; \citealt{GaiaCollaboration2023A&A}) came the non-single stars (NSS) catalog which contains astrometric orbital solutions for over 160,000 binaries \citep{GaiaCollaboration2023A&Ab}. This provided an unprecedentedly rich dataset from which several large samples of WDMS binaries have so far been compiled. While epoch astrometry of all sources used to construct the NSS catalog is not publicly available in DR3 which complicates the study of the selection function, \citet{El-Badry2024OJAp} showed that using a forward modeling approach starting with the Gaia scanning law, it is possible to tackle this challenge. For a binary with a given set of intrinsic parameters, their code, \texttt{gaiamock}\footnote{https://github.com/kareemelbadry/gaiamock/}, is able to simulate 1D astrometric measurements that would be obtained by Gaia and from these, predict whether or not it receives an orbital solution in the NSS catalog  and if so, provide the derived orbital parameters. Applying their code to a simulated population of Galactic sources, they show that they are able to reproduce the key features of the true NSS catalog, demonstrating that their method provides an accurate representation of the selection function. 

Recently, \citet{Yamaguchi2025PASP} used a detailed model of the Gaia selection function to simulate the sample of $\sim 3000$ WDMS binaries from \citet{Shahaf2024MNRAS}. This sample was selected based on astrometric orbits using a quantity called the ``astrometric mass ratio function" (AMRF; \citealt{Shahaf2019MNRAS}). This quantity can be used to select compact object binaries as systems whose orbits are too large at fixed period to be explained by any single nondegenerate companion. The majority of WDs in the sample have inferred masses near $\sim 0.6\,M_{\odot}$ \citep{Hallakoun2024}, suggesting they formed from an AGB progenitor. \citet{Yamaguchi2025PASP} could reproduce the broad demographics of the sample under the ansantz that MT from AGB donors remains stable if the accretor-to-donor mass ratio exceeds $0.4$. Additionally, they found that the intrinsic orbital period distribution of WDMS binaries is close to flat over $\sim100-1000\,$d and that $\sim0.4\%$ of all solar-type stars have a post-AGB WD companion in this period range.

A natural extension of the work by \citet{Yamaguchi2025PASP} is to model other observed samples of WDMS binaries within the same framework. Short-period post-CEE binaries, and wide WDMS binaries containing low-mass WDs formed via stable MT from RGB donors, are natural and complementary comparison datasets. A successful model of binary MT should be able to reproduce the properties of all observed samples, after accounting for each sample's selection function.

In this work, we extend the methods of \citet{Yamaguchi2025PASP} to model two additional samples of WDMS binaries: (1) a ``UV excess" sample of astrometric binaries from GALEX, which is sensitive to hot WDs and particularly systems containing low-mass helium WDs, and (2) an ``eclipsing binary'' (EB) sample of post-common envelope binaries (PCEBs) selected based on their positions on the Gaia color-magnitude diagram (CMD) and eclipses in ZTF light curves, which is sensitive to WDs orbited by M dwarfs in $\lesssim 2\,$d orbits \citep{Shariat2026PASP}. Based on the additional constraints provided by these samples, we also revisit the original ``AMRF sample" of \citet{Shahaf2024MNRAS}. Crucially, all three of these samples are selected homogeneously and have well-understood selection functions. %, which is necessary to make comparisons between the simulated and observed populations. 

The remainder of this paper is organized as follows. In Section \ref{sec:observed_samples}, we describe how the observed samples were selected and how we model this selection in our simulations. In Section \ref{sec:forward_model}, we describe our binary evolution model that takes simulated zero-age binaries and transforms them into present-day WDMS binaries. This includes the treatment of MT stability and orbital evolution through stable and unstable MT from donors at different evolutionary phases. We then describe how we recreate the selection of the observed samples in Section \ref{sec:model_selections}. In Section \ref{sec:results}, we present the results of our forward model. We compare properties of the observed and simulated samples, and study the features of the inferred intrinsic population of WDMS binaries. Finally, we conclude and summarize our key findings in Section \ref{sec:conclusions}. 

\section{Observed samples} \label{sec:observed_samples}

\begin{figure*}
    \centering
    \includegraphics[width=1.0\linewidth]{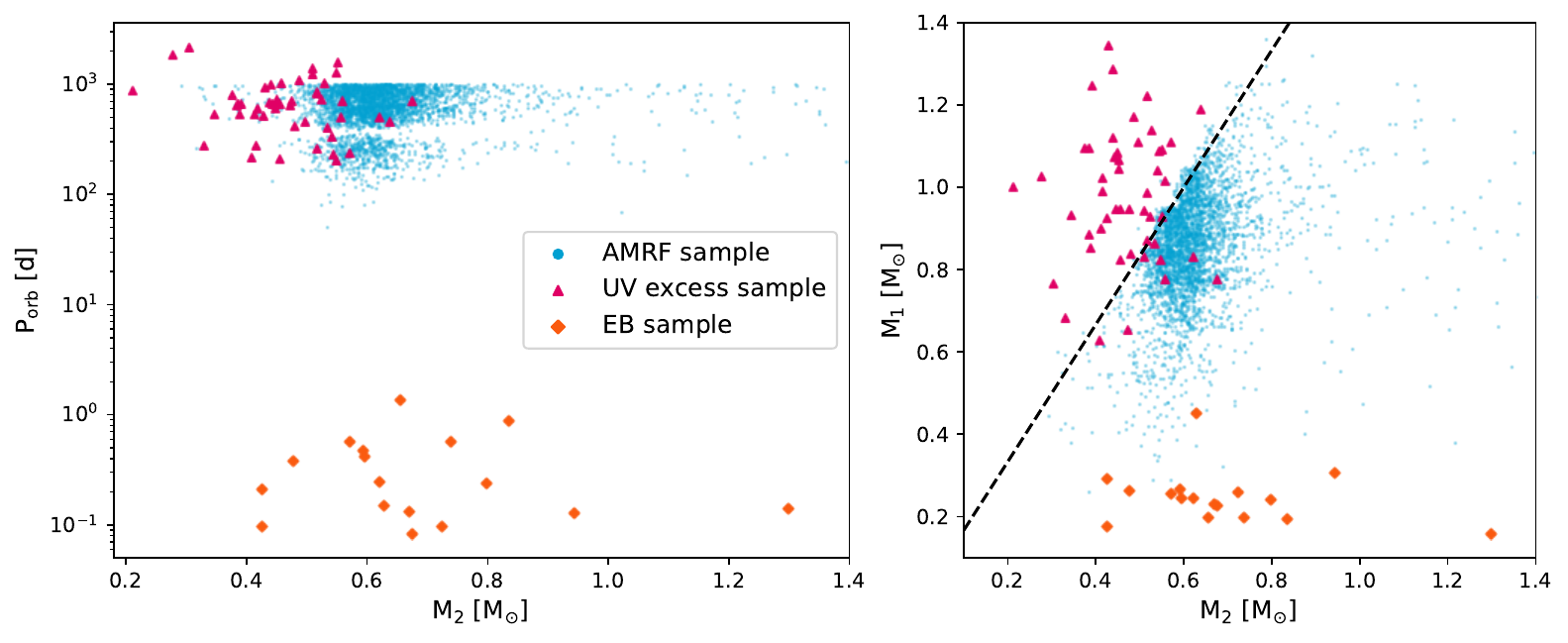}
    \caption{Orbital period (\textit{Left}) and companion star mass (\textit{Right}) against WD mass. We compare the three different observed samples considered in this work (Section \ref{sec:observed_samples}). The dashed line in the right panel marks a mass ratio of 0.6, the minimum mass ratio to which the AMRF sample is sensitive. The AMRF sample is dominated by CO WDs in AU-scale orbits, likely products of stable MT from AGB donors. Selected from the same parent Gaia catalog as the AMRF sample, the UV excess sample similarly probes AU-scale orbits but is sensitive to WDs of all masses, including He WDs with RGB progenitors. The EB sample is made up of short-period PCEBs hosting M dwarfs.}
    \label{fig:obs_samples}
\end{figure*}

Our model aims to reproduce three observed samples of WDMS binaries simultaneously. In the left panel of Figure \ref{fig:obs_samples}, we show orbital periods and WD masses of all systems. In the right panel, we show component masses. The three samples occupy distinct regions in these spaces. Below, we summarize how each of these samples were constructed. 

\subsection{AMRF sample} \label{ssec:amrf_observed}

\citet{Shahaf2024MNRAS} compiled a sample of 3145 high-probability WDMS binaries with astrometric orbital solutions from Gaia DR3. Their selection can be summarized in two steps: 
\begin{enumerate}
    \item For each astrometric binary hosting a MS primary, calculate the AMRF: 
    \begin{equation}
         \mathrm{AMRF} = \frac{\alpha}{\varpi} \left(\frac{M_1}{M_{\odot}} \right)^{-1/3} \left(\frac{P_{\rm orb}}{\rm yr} \right)^{-2/3}
    \end{equation}
    where $\alpha$ is the angular photocentric semi-major axis, $\varpi$ is the parallax, $P_{\rm orb}$ is the orbital period, and $M_1$ is the primary mass -- all observables which are measured by Gaia or, in the case of $M_1$, can be inferred from Gaia data. It can be shown that for any given $M_1$, there is a maximum possible AMRF for a MS secondary. Therefore, \citet{Shahaf2024MNRAS} define a boundary in AMRF-$M_1$ space above which systems are inconsistent with any MS secondary. These systems make up the so-called ``non-class I" (NCI) sample. 
    \item For each system in the NCI sample, calculate the difference between the observed $(B-I)$ color and that predicted for the MS primary. \citet{Shahaf2024MNRAS} discard systems with significant color excess (meaning they are redder than predicted) to remove possible triples hosting an inner binary of two low-mass MS stars. This leaves behind a final ``no color excess" (NCE) sample of astrometric binaries with compact object secondaries (most commonly, cool WDs).
\end{enumerate}
For more details on these steps, we refer readers to \citet{Shahaf2024MNRAS} and \citet{Yamaguchi2025PASP}. 

As shown in Figure \ref{fig:obs_samples}, the AMRF sample is restricted to systems with $M_{\rm 2}/M_1 \gtrsim 0.6$ and contains few WDs below $0.5\,M_{\odot}$. This is primarily due to the AMRF cut, which requires the secondary to be sufficiently massive that a MS companion can be excluded. 

\subsection{UV excess sample} \label{ssec:uv_observed}

The AMRF sample is insensitive to WDMS binaries with mass ratios below $\sim 0.6$, where single MS secondaries cannot be ruled out. This means it contains few systems with He WDs, which are products of stable MT from less evolved, RGB donors. Although low-mass WDs cannot be selected purely dynamically, they can be identified based on UV excess when they are young and hot. To select such systems, we cross-matched the DR3 astrometric binary sample with the GALEX catalog to identify systems with FUV excess, indicating the presence of a hot WD companion. We describe the basic procedure below: 
\begin{enumerate}
    \item Select astrometric binaries in the NSS catalog with \texttt{goodness\_of\_fit} $<10$ and GALEX FUV $< 17$. This gives us a starting set of 408 UV-bright astrometric binaries. Some of these contain WDs, but others simply contain hot MS stars. 
    \item Fit optical-to-infrared broadband photometry with a spectral energy density (SED) model for a single luminous star. We include Gaia G/BP/RP, synthetic SDSS g/r/i/z, WISE W1/W2/W3, and 2MASS J/H/Ks. For the fitting, we use the \texttt{MINEsweeper} code to interpolate MIST evolutionary models and generate synthetic photometry. The free parameters are initial stellar mass, [Fe/H], parallax, and equivalent evolutionary phase (EEP; a quantity that increases monotonically with age). We select candidates with a GALEX FUV magnitude at least 2 magnitudes brighter than the best-fit SED model prediction. We also remove sources with extinction-corrected $(BP-RP)<0.5$, as we found that many such candidates are simply UV-bright A and B stars with imperfect SED fits. This leaves us with 79 binaries. 
    \item Carry out two-component SED fitting with the VO SED Analyzer (VOSA; \citealt{Bayo2008A&A}) tool. For the WD, we use Koester WD models with hydrogen atmospheres \citep{Koester2010MmSAI} and for the MS star, we use ATLAS9 Kurucz ODFNEW / NOVER models \citep{Castelli2003IAUS}. We then remove sources with an inferred radius of $>0.5\,R_{\odot}$ for the hotter component to further exclude possible MS stars. This gives us a total of 63 binaries in the final sample. 
\end{enumerate}
As part of a Cycle 33 SNAP program (PID 18120), we are obtaining UV spectra for a random subset of these binaries using the Cosmic Origins Spectrograph (COS) on the Hubble Space Telescope. A more detailed description of the selection summarized above, along with an analysis of the COS spectra, will be presented in future work (Yamaguchi et al, in prep). 

The right panel in Figure \ref{fig:obs_samples} shows that the UV excess sample probes $M_2/M_1\lesssim 0.6$, which is inaccessible to the AMRF sample. While less massive WDs have larger radii, which makes them brighter at a given temperature, more massive WDs cool more slowly and thus remain UV-bright longer. These competing effects cause the UV excess sample to provide a less biased view of the underlying WD mass distribution than the AMRF sample (Section \ref{sec:results}). %We note that WD masses for both samples were derived from their astrometric orbits and primary mass estimates, assuming that the WDs contribute negligibly to the total G-band flux. 

\subsection{EB sample} \label{ssec:eb_observed}

Both the AMRF and UV excess samples are made up of astrometric binaries from Gaia DR3. This means that they are limited to orbital periods of $\sim100-1000\,$d, corresponding to AU-scale separations. A different sample, not selected by astrometry, is required to probe the shorter periods expected for post-common envelope binaries (PCEBs). 

Recently, \citet{Shariat2026PASP} (hereafter \citetalias{Shariat2026PASP}) compiled a sample of short-period ($P_{\rm orb}\lesssim2\,$d) detached eclipsing WDMS binaries, using simple cuts based on CMD position and detected eclipses in ZTF light curves. In brief, they select systems that lie between the MS and WD cooling tracks on the Gaia CMD that are in the northern sky and within 200 pc of Earth. After some additional quality cuts, they were left with 3777 CMD-selected WDMS candidates, from which they identified 39 detached eclipsing binaries via an automated period search of ZTF light curves and manual vetting of the final candidates. The selection function of the sample is well-understood. The CMD selection is made up of simple cuts, and \citetalias{Shariat2026PASP} conducted extensive injection-recovery tests to quantify the ZTF selection function, and using these, they inferred completeness-corrected period and companion mass distributions. Because all the systems in this sample have short periods and host low-mass MS stars, they are almost certainly PCEBs.

\section{Forward model} \label{sec:forward_model}

\subsection{Initial binary population} \label{ssec:galaxia}

We initialize with a realistic Galactic stellar population, whose positions, metallicities, and kinematics are generated using the code \texttt{Galaxia} \citep{Sharma2011ApJ}. Ages are drawn uniformly between $0$ and $12\,$Gyr. A fraction of these stars are assigned binary companions using \texttt{COSMIC} \citep{Breivik2020ApJ}, which draws orbital parameters from distributions inferred by \citet{Moe2017ApJS}. Additionally, a fraction of the binaries are assigned outer tertiaries based on the multiplicity fraction as a function of primary mass from \citet{Offner2023ASPC}. Our model only considers a tertiary as an extra source of light close to the binary that may or may not affect the total observed flux and astrometric orbit of the system. We do not consider how these tertiaries affect the evolution of the inner binaries, which can be important \citep[e.g.][]{Naoz2013MNRAS, Naoz2016ARA&A, Toonen2016ComAC, Toonen2020A&A, Shariat2023ApJL} and may explain existing discrepancies between our simulated and observed samples (Section \ref{sec:results}), possibly motivating future extensions to our model to include effects of triple dynamics. 

We obtain synthetic photometry of each stellar component via interpolation of MIST isochrones \citep{Choi2016ApJ} using the \texttt{isochrones} package \citep{Morton2015ascl}. We calculate $E(B-V)$ extinctions using the \texttt{mwdust} package \citep{Bovy2016ApJ}, which combines dust maps from \citet{Drimmel2003A&A}, \citet{Marshall2006A&A}, and \citet{Green2019ApJ}. A more thorough description of the initial population can be found in Section 3 of \citet{El-Badry2024OJAp} and Section 2 of \citet{Yamaguchi2025PASP}. 

\subsection{Interaction and mass transfer} \label{ssec:MT_models}

Here we describe how the stellar and orbital properties of the initial MSMS binaries are evolved until the present day, with particular focus on systems that end up as post-MT WDMS binaries. 

\subsubsection{Donor evolutionary stage} \label{sssec:donor_evol}

First, we identify binaries in which the initial primary evolves off the MS and forms a WD. %This is done by interpolating MIST evolutionary tracks. A similar approach is described in Section 5.1 and 5.2 of \citet{Yamaguchi2025PASP}, although the interpolation method has since been improved. 
We carry out bilinear interpolation on mass and metallicity using MIST single-star evolutionary tracks \citep{Choi2016ApJ} to identify when, if at all, the radius of the primary star exceeds the Roche lobe radius at periastron ($r_{L, \rm peri}$) of the initial orbit. The equivalent evolutionary phase (EEP) at the time of interaction determines if the primary was an RGB ($454 < \mathrm{EEP}\leq707$) or an AGB ($707<\mathrm{EEP}\leq1409$) donor. This interpolation also returns the mass of the star at this time (donor mass; $M_d$) and its core mass ($M_c$). $M_d$ is roughly equal to the initial mass on the RGB but can be significantly lower than the initial mass on the AGB due to wind mass loss.

\subsubsection{Stability boundary} \label{sssec:stability}

% AGB 

For systems that interact with AGB donors, following \citet{Yamaguchi2025PASP}, we set a constant critical mass ratio, $q_{\rm crit, AGB} = M_a/M_d = 0.4$. Systems with mass ratios smaller (more unequal) than this are modeled as undergoing unstable mass transfer (Section \ref{sssec:cee}). The value of $q_{\rm crit, AGB}$ is set primarily by the number of WDMS binaries that enter the astrometric sample.

% RGB 

In the case of RGB donors, we interpolate the quasi-adiabatic critical mass ratios ($q_{\rm qad}$) from \citet{Temmink2023AA} on $M_d$ and $R_d$. We also set a fixed lower limit on this value, $q_{\rm crit, RGB, min}$, which translates to a stricter requirement for stability (i.e. $q_{\rm crit, RGB} =\mathrm{max}[q_{\rm qad}, q_{\rm crit, RGB, min}]$) and which we set to a fiducial value of $0.65$. We do not employ the same critical mass ratio as for the AGB donors, as doing so results in too many low-mass He WDs which enter the simulated AMRF sample (Figure \ref{fig:no_qcritminRGB}, Section \ref{sssec:comparison_amrf}). We identify sub-giant branch (SGB) donors using the radius-mass relation at the base of the RGB determined by \citet{Temmink2023AA} (See their Figure 8). We use $q_{\rm qad}$ for the critical mass ratios for SGB donors. 

% MS

If RLOF occurs on the MS, we assume that the system merges and remove it from the population. %Fewer than 0.01\% of interacting systems fall under this category. 
While some such systems will be observable before merger as contact binaries, they are unlikely to produce WDMS binaries that enter our samples.

\begin{figure}
    \centering
    \includegraphics[width=1.0\linewidth]{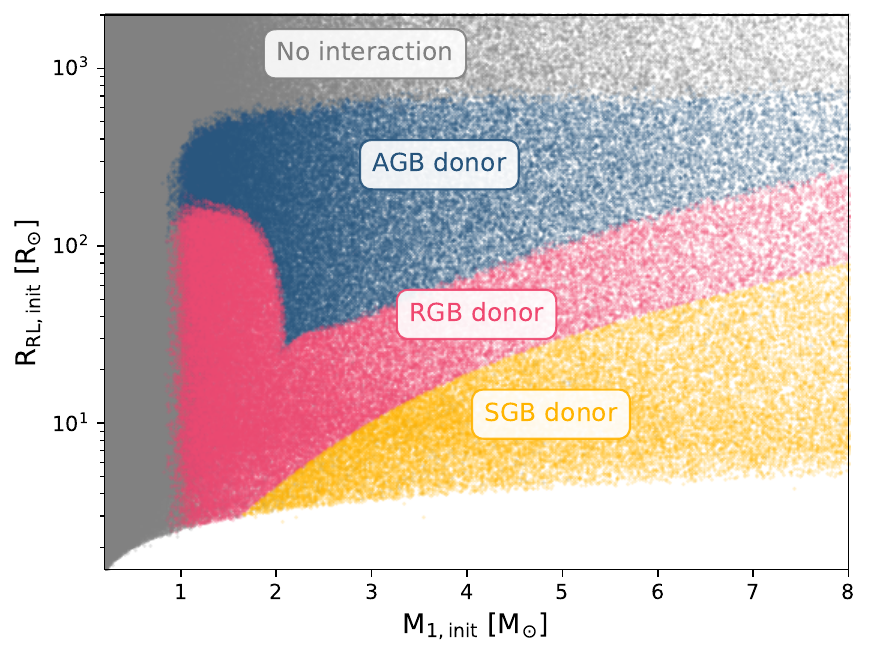}
    \caption{Initial Roche lobe radius against initial primary mass for a subset of our simulated systems. Points are colored by the evolutionary phase of the primary when its radius reaches the Roche lobe radius. If this never happens, we classify the system as undergoing no interaction (gray), and it experiences no change in orbital separation.}
    \label{fig:interaction}
\end{figure}

In Figure \ref{fig:interaction}, we show the initial Roche lobe radius at pericenter, $R_{\rm RL, init}$, and initial primary mass, $M_{\rm 1,init}$, for simulated binaries which interact with donors in different evolutionary stages. Note that for interacting systems, the initial primary ultimately becomes the WD, which is the photometric secondary today. Systems with $M_{\rm 1,init} \lesssim 1\,M_{\odot}$ do not have enough time to evolve off the MS and expand to undergo interaction in the age of the Universe. Systems with $R_{\rm RL, init}\gtrsim 10^3\,R_{\odot}$ also do not interact as the primaries never overflow their Roche lobes. The boundary between RGB and AGB donors is predicted by the MIST evolutionary tracks (Section \ref{sssec:donor_evol}) while the boundary between SGB and RGB donors is taken directly from \citet{Temmink2023AA}. 

\subsubsection{Stable MT from AGB donors} \label{sssec:stable_mt_agb}

For systems that experience stable MT when the donor is on the AGB, we determine the change in semi-major axis using the formalism from \citet{Soberman1997A&A}. By default, we assume fully non-conservative mass transfer with mass loss through isotropic re-emission from the accretor ($\beta$ = 1). For the WD mass, we follow \citet{El-Badry2024OJAp} and use the linear initial-final mass relation (IFMR) from \citet{Weidemann2000AA}, which is calibrated to observations of single WDs (also see Section 2.2 and Appendix A.4 of \citealt{Yamaguchi2025PASP}). To match the eccentricity distribution of the observed sample, which is peaked at $e\sim0.1$ (Section \ref{sssec:comparison_amrf}), we allow systems to retain some of their initial eccentricities. Following \citealt{Yamaguchi2025PASP}, we employ the following relation with an arbitrarily chosen functional form: $e_{\rm final}= 0.8e_{\rm init}^4 + 0.04$, to which we add Gaussian noise with a standard deviation of $0.05$. While this is a minimal treatment that does not encode any physics, several recent works show MT that occur in eccentric orbits can lead to residual eccentricities of comparable magnitudes \citep{Parkosidis2026A&A_a, Parkosidis2026A&A_b, Parkosidis2026arXiv_c}. In Appendix \ref{appendix:ecc}, we show results assuming no change from initial eccentricities (drawn from the thermal distribution of \citet{Moe2017ApJS}) as well as complete circularization (i.e., zero post-MT eccentricity), which both lead to worse agreement between the final eccentricity distributions of the simulated and observed samples. 

\subsubsection{Stable MT from RGB and SGB donors} \label{sssec:stable_mt_rgb}

For stable MT on the RGB, we again predict orbital evolution using the relations from \citet{Soberman1997A&A} with $\beta = 1$. We also implement the same relation between the initial and final eccentricity as for the AGB donors. However, we modify the WD IFMR so that the systems ultimately fall on the $P_{\rm orb}-M_{\rm WD}$ relation first derived in \citet{Rappaport1995MNRAS}. We draw $M_{\rm WD}$ from an asymmetric Gaussian capturing the factor of 2.4 uncertainty above and below the best-fit predicted relation as quoted in \citet{Rappaport1995MNRAS}. This relation has been shown to explain the orbits of millisecond pulsar + helium WD binaries, which are thought to be stable MT products from RGB donors \citep[e.g.][]{Corongiu2012ApJ, Hui2018ApJ}. 

For SGB donors, we assume the same orbital evolution but set WD masses to be the core mass at the onset of MT, $M_c$, and assume circularization. 

\subsubsection{Common envelope evolution} \label{sssec:cee}

For systems that experience unstable MT and undergo common envelope evolution, we use the $\alpha\lambda$-formalism to predict the amount of orbital shrinkage \citep{Livio1988ApJ, deKool1990ApJ}. This is a prescription based on energy conservation, where the two parameters are $\lambda$, which parameterizes the structure of the donor's envelope, and $\alpha$, the fraction of liberated orbital energy that goes into unbinding this envelope. It is the product of these parameters, $\alpha \lambda$, that determines the final separation. The ratio of the final to initial semi-major axis is given by:
\begin{equation} \label{eqn:alpha_lambda}
    \frac{a_f}{a_i} = \frac{M_{\rm WD}}{M_d}\left(1 + \frac{2M_{d, \rm env}a_i}{M_{a}R_{d}\alpha\lambda} \right)^{-1}
\end{equation} 
where $M_{d, \rm  env}$ is the donor envelope mass (which we take to be $M_d-M_{\rm WD}$), $R_d$ is the initial donor radius ($=r_{L, \rm peri}$), and $M_{a}$ is the accretor mass (which is equivalent to the present-day MS mass under the fully non-conservative assumption). The WD mass is set to be the smaller of $M_c$ and the prediction from \citet{Weidemann2000AA}. 

The simple analytic form of the $\alpha\lambda$-prescription has led to its extensive use in binary population synthesis codes \citep[e.g.][]{Breivik2020ApJ, Riley2022ApJS}. Previous studies modeling observed samples of short-period PCEBs have inferred a ``universal" value of $\alpha\lambda\sim0.2-0.3$ \citep[e.g.][]{Zorotovic2010A&A, Davis2012MNRAS, Scherbak2023MNRAS, Shi2026arXiv}. On the other hand, several works have highlighted limitations of this prescription, in particular with regard to double white dwarf binaries \citep{Nelemans2000A&A, Nelemans2005MNRAS} as well as several WDMS binaries which are possible PCEBs in $\sim 10-20\,$d orbits \citep{Wonnacott1993MNRAS, Lin2026ApJ, Yamaguchi2024MNRAS, Motherway2026AJ, Torres2025A&A}. For the purposes of this work, we use the prescription with a single, fixed value of $\alpha\lambda = 0.3$, as the PCEBs in the \citetalias{Shariat2026PASP} sample are limited to $P_{\rm orb} \lesssim 2\,$d. %In the future, with larger homogeneously selected samples at $\sim 10-20\,$d periods, this aspect of the model could be made more flexible. 

\subsubsection{Magnetic braking} \label{sssec:MB}

Stellar winds coupled to magnetic fields remove angular momentum from a star, causing the star to spin down over time \citep[e.g.][]{Schatzman1962AnAp, Weber1967ApJ, Skumanich1972ApJ}. However, for a star in a close binary, tides synchronize the orbital and rotation period, preventing spin-down \citep[e.g.][]{Zahn1977A&A}. In this case, angular momentum is removed from the orbit, leading to orbital shrinkage. This process is called magnetic braking (MB; e.g. \citealt{Verbunt1981A&A, Rappaport1983ApJ}). 

For WDMS binaries with post-MT periods $<20\,$d, we include the effect of MB on the post-MT period evolution. We use a ``saturated" MB model \citep{Kawaler1988ApJ, Chaboyer1995ApJ, Sills2000ApJ}, which predicts that the widely used $P_{\rm orb}^{-3}$ scaling of the MB torque from \citet{Rappaport1983ApJ} does not hold down to arbitrarily fast rotation rates, and instead saturates beyond some critical value. Several observational works have since found evidence of this behavior \citep[e.g.][]{Reiners2009ApJ, El-Badry2022MNRAS}. Following \citet{Belloni2024A&A}, we introduce two free parameters, $k$ and $\eta$, to the saturated MB torque, $\dot{J}_{\rm sat}$: 
\begin{equation}
    \dot{J}=
    \begin{cases}
    k\dot{J}_{\rm sat} & \text{if } M_{\rm 1}> M_{\rm conv}\\
    \frac{k}{\eta}\dot{J}_{\rm sat} & \text{if } M_{\rm 1}< M_{\rm conv}
    \end{cases}
\end{equation}
where $k > 1$ ``boosts" the strength of MB from that calibrated to single MS stars, and $\eta>1$ ``disrupts" (weakens) MB below a critical mass $M_{\rm conv} \sim 0.35\,M_{\odot}$, which corresponds to the fully convective boundary \citep[e.g.][]{Chabrier1997A&A, Baraffe1998A&A}. Observed samples of post-interaction binaries hosting WDs and hot subdwarfs with MS companions have found that a saturated MB model can explain their period and companion mass distributions, but only with strong disruption and boosting ($\eta, k\gtrsim50$; e.g. \citealt{Schreiber2010A&A, Belloni2024A&A, Blomberg2024PASP}). By default, we set $k = \eta = 50$, but we discuss the effect of changing these values on the short-period EB sample in Section \ref{sec:results}. The analytical solution for the orbital period evolution is given in Appendix D.2 of \citet{Blomberg2024PASP}. 

%Given the present-day periods after MB losses, we remove systems where the MS star overfills the Roche lobe, becoming a cataclysmic variable. 
We remove systems in which the MS star fills its Roche lobe and becomes a cataclysmic variable since the \citetalias{Shariat2026PASP} sample only contains detached systems.

\subsubsection{WD cooling models} \label{sssec:wd_cooling}

To predict brightnesses of the WDs in the Gaia G and GALEX UV bands, we use cooling models from \citet{Bergeron2020} \footnote{https://www.astro.umontreal.ca/~bergeron/CoolingModels/}. The calculation of synthetic photometry is detailed in \citet{Holberg2006AJ}. We use extinction coefficients from \citet{Yuan2013MNRAS} to convert $E(B-V)$ to extinctions in the GALEX passbands. 

By default, for WDs below $0.6\,M_{\odot}$, we use the ``thick" H atmosphere models with H mass fraction $q_H=10^{-4}$. Above $0.6\,M_{\odot}$, we use ``thin" H atmosphere models with $q_H=10^{-10}$, which cool faster. We find that this better reproduces the WD mass distribution of systems in the UV excess sample than using either thick or thin atmosphere models over all masses (Section \ref{sec:results}). Theoretical studies predict that the thickness of a WD's hydrogen layer decreases with increasing mass, consistent with our assumptions, though the difficulty of making accurate measurements of envelope mass makes observational confirmation of this challenging \citep[e.g.][]{Romero2012MNRAS, Romero2019MNRAS, Bauer2026ApJS, Arseneau2026arXiv}

\section{Selection functions} \label{sec:model_selections}

In this section, we describe our process to reproduce the cuts made to select the observed samples described in Section \ref{sec:observed_samples}. 

\subsection{Mock AMRF sample} \label{ssec:model_sel_amrf}

For details of the steps to construct a mock AMRF sample following \citet{Shahaf2024MNRAS}, we refer readers to \citet{Yamaguchi2025PASP}. In brief, we simulate Gaia observations of each binary using gaiamock and calculate the AMRF to isolate systems that would enter the NCI sample. We then interpolate PARSEC models (version 1.2S; \citealt{Chen2014MNRAS, Tang2014MNRAS, Chen2015MNRAS}) using the \texttt{stam} package \citep{Hallakoun2021MNRAS}\footnote{https://github.com/naamach/stam} to obtain $(B-I)$ colors that would be observed given the true system parameters and those that would be predicted following the approach of \citet{Shahaf2024MNRAS}. Using these, we calculate color excess for each binary. We retain objects without significant color excess to construct a mock NCE sample.

\citet{Yamaguchi2025PASP} showed that color excess values calculated as described above fail to reproduce the observed trend with primary mass and to address this, implemented a purely empirical mass-dependent correction to the color excess.
% The discrepancy was originally attributed to the fixed age assumption, where solar-type stars appear bluer as they evolve on the MS leading to bluer than expected colors if ages are underestimated, as well as to possible inaccuracies in the colors predicted by the PARSEC models. To tackle this, 
%\citet{Yamaguchi2025PASP} implemented a purely empirical correction to the color excess. %CE as a function of primary mass by fitting the observed data points.
We have since determined that the primary source of this discrepancy is a systematic and mass-dependent offset in the metallicities of the observed systems calculated by \citet{Zhang2023MNRAS} and used by \citet{Shahaf2024MNRAS} when calculating color excess. By calibrating to metallicities from APOGEE \citep{Majewski2017AJ, Nidever2015AJ, Wilson2019PASP}, we derive a fit to the offset of \citet{Zhang2023MNRAS} metallicities as a function of absolute magnitude, which we apply to the simulated metallicities used to predict expected colors. This allows us to reproduce the general trend in color excess vs. primary mass and, with an additional constant offset in color excess, the normalization. Comparing to the results of \citet{Yamaguchi2025PASP}, the change in the color excess calculation does not significantly affect the inferred intrinsic parameter distributions, but does increase the predicted contribution from contaminating triples. We describe this process in detail in Appendix \ref{appendix:ce_calc}. 

\subsection{Mock UV excess sample}

For the UV excess sample, we select systems with synthetic GALEX FUV $<17$. We estimate the FUV excess as the difference between the total and MS-only FUV magnitudes. We use maps from the CDS MOS Server\footnote{https://aladin.cds.unistra.fr/hips/list} to isolate systems which are in the GALEX footprint, and remove simulated systems exceeding the GALEX saturation limit with NUV and FUV magnitudes brighter than $8.9$ and $9.5$. We also remove systems with predicted total $(BP-RP) < 0.5$, as done for the observed UV excess sample. 

\subsection{Mock EB sample}

\begin{figure*}
    \centering
    \includegraphics[width=1.0\linewidth]{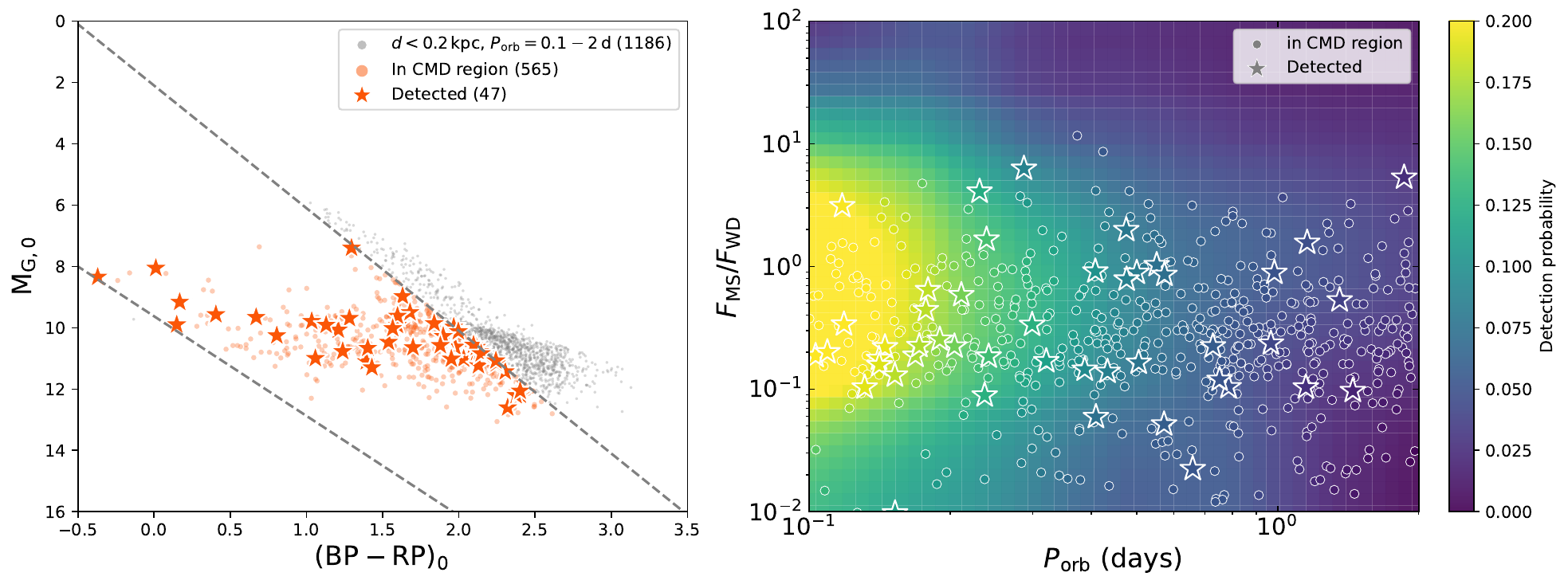}
    \caption{\textit{Left}: All simulated WDMS binaries with $P_{\rm orb} = 0.1- 2\,$d within $0.2\,$kpc on the Gaia CMD. The dashed lines define the boundary between the MS and WD cooling track used to select WDMS binaries in \citetalias{Shariat2026PASP}. Systems lying in the region defined by these boundaries are shown in orange. In star markers, we show one instance of the final detected sample accounting for the detection probability via eclipses in ZTF. \textit{Right}: The detection probability as a function of G-band flux ratio and orbital period for systems that lie in the selected CMD region.}
    \label{fig:cmd_ebs}
\end{figure*}

For the EB sample, we select WDMS binaries between the MS and the WD cooling track, as done by \citetalias{Shariat2026PASP}. In the left panel of Figure \ref{fig:cmd_ebs}, we plot simulated WDMS binaries with $P_{\rm orb} =0.1-2\,$d on the Gaia CMD. As \citetalias{Shariat2026PASP} manually inspected all light curves, we assume that their sample of 39 eclipsing binaries is pure. In total, 565 WDMS binaries with $P_{\rm orb} =0.1-2\,$d lie in the selected CMD region within 200$\,$pc. Only a fraction of these will show visible eclipses in ZTF light curves. We calculate the detection probability as a function of the G-band flux ratio and orbital period, based on the grid of synthetic light curves generated by \citetalias{Shariat2026PASP}, which we plot in the right panel of Figure \ref{fig:cmd_ebs}. In \citetalias{Shariat2026PASP}, this is denoted as $f_{\rm detected\mid CMD} \times f_{\rm in\ ZTF}$, where the first factor incorporates all ZTF systematics (cadence, noise, etc.) and the second is a constant value of $2/3$ corresponding to the ZTF sky coverage. We use this to randomly select the final EB sample -- one such instance is plotted with star markers on the left panel. The detection probability is highest for short-period systems in which the WD and MS star contribute similar fractions of the G-band flux. It is always significantly lower than 1, since most systems are not sufficiently edge-on to show eclipses. 

\section{Results} \label{sec:results} 

\subsection{Comparison to observed samples} \label{ssec:comparisons}

\subsubsection{AMRF sample} \label{sssec:comparison_amrf}

\begin{figure*}
    \centering
    \includegraphics[width=1.0\linewidth]{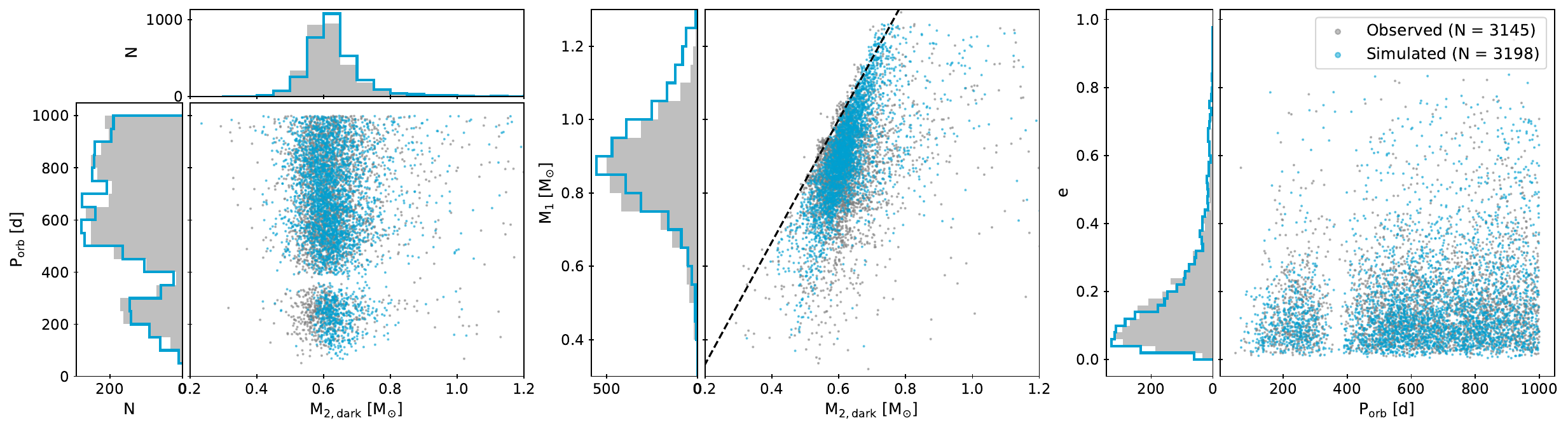}
    \caption{Distributions of several stellar and orbital parameters of the observed (gray) and simulated (blue) AMRF samples. The black dashed line in the central panel marks a mass ratio of 0.6. Overall, our model reproduces most key features of the observed sample (e.g., total number, narrow $M_{\rm WD}$ distribution, trends in $P_{\rm orb}$, non-zero eccentricities), though there remain a few tensions (i.e., a stronger than observed correlation between component masses, and slightly more massive MS stars).  }
    \label{fig:amrf_results}
\end{figure*}

In Figure \ref{fig:amrf_results}, we compare distributions of several orbital and stellar parameters of systems in the simulated and observed AMRF samples. Overall, we find good agreement between both the 1D and 2D distributions and in the total number of simulated and observed systems. In particular, the simulated orbital period distribution reproduces the observed distribution, while the WD mass distribution is sharply peaked at $\sim 0.6\,M_{\odot}$, as observed. These results are almost identical to those of \citet{Yamaguchi2025PASP} since we have made minimal changes to the assumed evolution via stable MT from AGB donors. However, in their work, they neglected products of stable MT from RGB donors. 

\begin{figure*}
    \centering
    \includegraphics[width=0.95\linewidth]{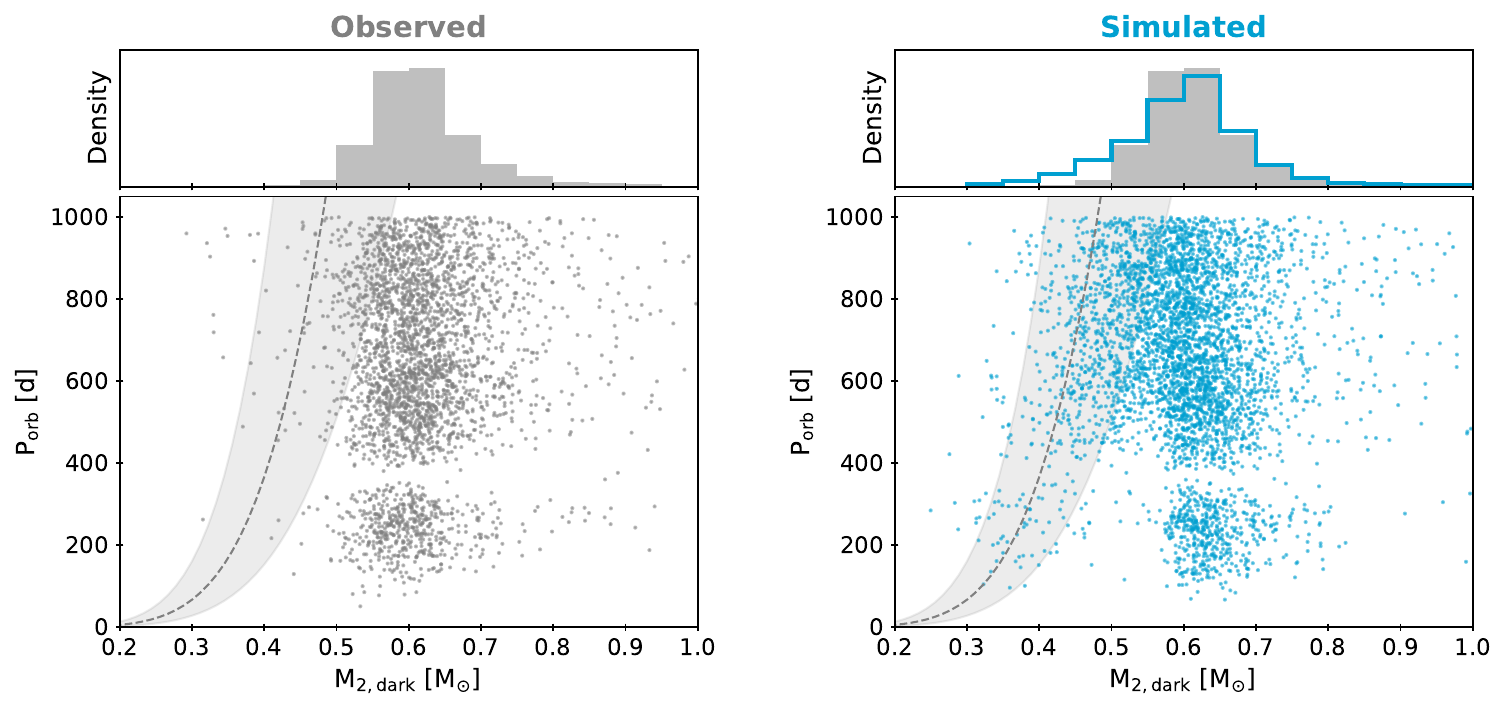}
    \caption{Similar to the leftmost panel of Figure \ref{fig:amrf_results}, but with the observed sample alone (\textit{Left}) and the simulated sample that results from imposing no lower limit on $q_{\rm crit,RGB}$ from \citet{Temmink2023AA} (\textit{Right}). The histograms have been normalized to isolate the effect of the increased contribution from post-RGB MT products, which host low-mass WDs with $\lesssim 0.5\,M_{\odot}$ and which lie near the predicted relation from \citet{Rappaport1995MNRAS} (gray shaded region). Adopting $q_{\rm crit,RGB}$ from \citet{Temmink2023AA}, or using the same value used for AGB donors, produces too many predicted low-mass WDs with short periods.   }
    \label{fig:no_qcritminRGB}
\end{figure*}

In this work, we instead use critical mass ratios for RGB donors from \citet{Temmink2023AA}, but set a lower bound, $q_{\rm crit, RGB, min}$, of $0.65$ (Section \ref{sssec:stability}). On average, this means that accretors (i.e., the present-day MS star) must be more massive to maintain stability, translating to lower final secondary-to-primary mass ratios, which the AMRF cut is biased against. In Figure \ref{fig:no_qcritminRGB}, we plot $P_{\rm orb}-M_{\rm WD}$ of the observed and simulated sample without $q_{\rm crit, RGB, min}$, where we see that there are products of stable MT from RGB donors hosting low-mass WDs which are not present in the observed sample. Note that this does not largely affect the number that enters the UV excess sample, since the majority of all systems that interact on the RGB satisfy this cut. 

As found in \citet{Yamaguchi2025PASP}, we find that the simulated primary masses are skewed to higher values than observed, and the predicted correlation between primary mass and WD mass is stronger than observed. This may point to our input mass ratio distribution being too top-heavy, and/or to our empirical model with simplified assumptions about AGB MT (e.g., a fixed critical mass ratio on the AGB, a single star WD IFMR) not capturing all of the relevant physics.

Lastly, by construction of the initial-to-final eccentricity relation (Section \ref{sssec:stable_mt_agb}), the observed and simulated distributions of eccentricities are in close agreement with each other. Once again, this suggests that tidal circularization is inefficient in these binaries and/or there are important eccentricity pumping mechanisms pre- or post-MT. The latter includes interaction with an outer tertiary or a circumbinary disk (see \citealt{Naoz2016ARA&A, Lai2023ARA&A} respectively for recent reviews on these topics). We mention that several other distinct populations of post-MT binaries, including those hosting blue straggler stars and barium stars, have been identified to have non-zero eccentricities, the origins of which also remain unclear \citep[e.g.,][]{Mathieu2025ARA&A, vanSon2026arXiv}.

\subsubsection{UV excess sample} \label{sssec:comparison_uv_excess}

\begin{figure*}
    \centering
    \includegraphics[width=1.0\linewidth]{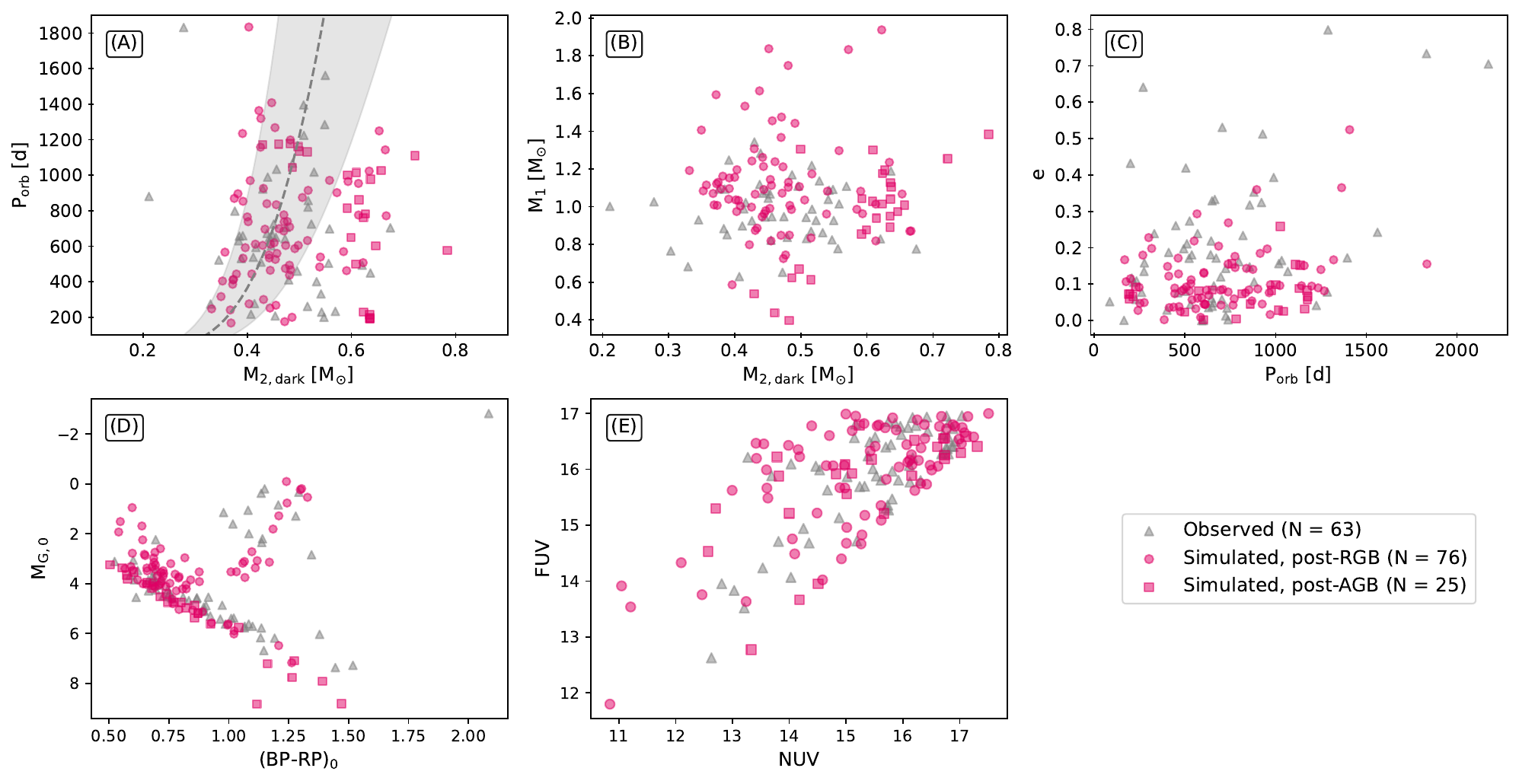}
    \caption{Fitted parameters and photometry of the observed (gray) and simulated (pink) UV excess samples. Post-RGB and post-AGB MT products are distinguished by their markers. The dashed line in panel A is the theoretical relation for post-RGB stable MT products derived by \citet{Rappaport1995MNRAS} with the reported uncertainty shaded in gray.}
    \label{fig:uvexcess_results}
\end{figure*}

We similarly compare the simulated and observed UV excess samples in Figure \ref{fig:uvexcess_results}. Once again, the two samples are in reasonable, but not perfect, agreement. While several studies report offsets from the $P_{\rm orb}-M_{\rm WD}$ relation of \citet{Rappaport1995MNRAS} \citep[e.g.][]{Tauris1999A&A, Nelson2004ApJ, Gao2023MNRAS}, we find that the spread in the observed data is well reproduced by this relation and the quoted uncertainty (panel A). Additionally, there are both observed and simulated systems that lie rightward of the relation, which can be attributed to post-AGB stable MT products that also enter this sample. 

In panel B, we see that the simulated sample contains more massive primaries, with 11 systems having $M_1 > 1.4\,M_{\odot}$ where there are none observed. However, we mention that observed systems with giant primaries have not been plotted, as these do not receive mass estimates in the \texttt{binary\_masses} table \citep{GaiaCollaboration2023A&Ab}. This accounts for 4 out of the 11 simulated systems. Meanwhile, the remaining 7 are not yet on the giant branch but are near the end of their MS phase, with $M_{\rm G,0} \lesssim 3$, where there is only one such observed system. About half of these may not enter the \texttt{binary\_masses} table, and therefore, the discrepancy seen in the $M_1$ distribution is likely exaggerated. However, why there are so few observed systems in the same region of the CMD, near terminal-age MS, remains unclear. 

Compared to the observed sample, there are also more massive WDs ($\gtrsim 0.6\,M_{\odot}$) with AGB progenitors in the simulated sample. This number is however sensitive to the cooling models. For this very reason, we used models with ``thin" H envelopes for WDs above $0.6\,M_{\odot}$ which cool faster than those with ``thick" H envelopes (Section \ref{sssec:wd_cooling}). Despite this, it is possible these models are still underestimating cooling rates at the high-mass end. Another possible factor is the WD IFMR used for AGB progenitors from \citep{Weidemann2000AA}, which was calibrated using single stars. Therefore, it does not account for the possible truncation of core growth at the onset of MT and could lead to overestimated WD masses. However, a significant deviation from this IFMR would be inconsistent with  the fact that we are able to reproduce the WD mass distribution and total number in the AMRF sample (Section \ref{sssec:comparison_amrf}). Additionally, there are a few extremely low-mass (ELM) WDs $\lesssim 0.3\,M_{\odot}$ in the observed sample where there are none predicted. We note that some of the observed systems may have poor orbital solutions or inaccurate UV photometry. In the near future, UV spectroscopy will confirm or rule out the presence of a WD for a subset of these systems, while multi-epoch radial velocity follow-up in the optical will allow us to confirm the Gaia orbital solutions. 

In panel C, we see that the observed eccentricities are, on average, higher than those of the simulated sample. This is despite the fact that the same initial-to-final eccentricity relation is able to explain the observed eccentricities in the AMRF sample (Section \ref{sssec:comparison_amrf}). This suggests that post-RGB MT WDMS binaries remain in significantly eccentric orbits, possibly even more so than the post-AGB MT systems that dominate the AMRF sample. This is in contrast to the expectation from binaries hosting millisecond pulsars around He WDs, the majority of which are in highly circular orbits with $e \lesssim 10^{-3}$, consistent with the prediction from \citet{1992RSPTA.341...39P}. Eccentric orbits in post-RGB systems have also recently been pointed out by \citet{Moltzer2025A&A}. We note that the three most eccentric systems with $e \gtrsim 0.7$ have orbital periods near the upper limit for Gaia DR3. While their reported \texttt{goodness\_of\_fit} values are not systematically higher than other systems in the sample, which all have values $< 10$, it remains a possibility that some systems have unreliable orbits. Once again, radial velocities will be important to validate their orbits and study the eccentricities of the UV excess sample in more detail.  

From panel D, our model predicts a similar number of evolved primaries as are found in the observed sample. However, the simulated population contains more sources redward of the main sequence; these are sources with $\gtrsim 1\,M_{\odot}$ primaries near the end of their main-sequence evolution. This ties into the previous discussion of panel B, where we predict more systems with $M_1 \gtrsim 1.4\,M_{\odot}$ than are observed. 
%In theory, we expect our sample to be biased against these primaries as they are both bright compared to the unevolved stars and blue compared to the giants, meaning that the WDs must be hotter to observe the same UV excess. It is noted that these slightly evolved primaries are also the most massive, making them responsible for the tail at $M_1 \gtrsim 1.75\,M_{\odot}$.

Finally, panel E shows that the simulations produce a distribution of NUV and FUV magnitudes similar to what is observed, indicating that the synthetic photometry predicted by the WD cooling tracks is reasonable. 

% There are slightly more total number - more evolved systems

\subsubsection{EB sample} \label{sssec:comparison_eb}

\begin{figure*}
    \centering
    \includegraphics[width=1.0\linewidth]{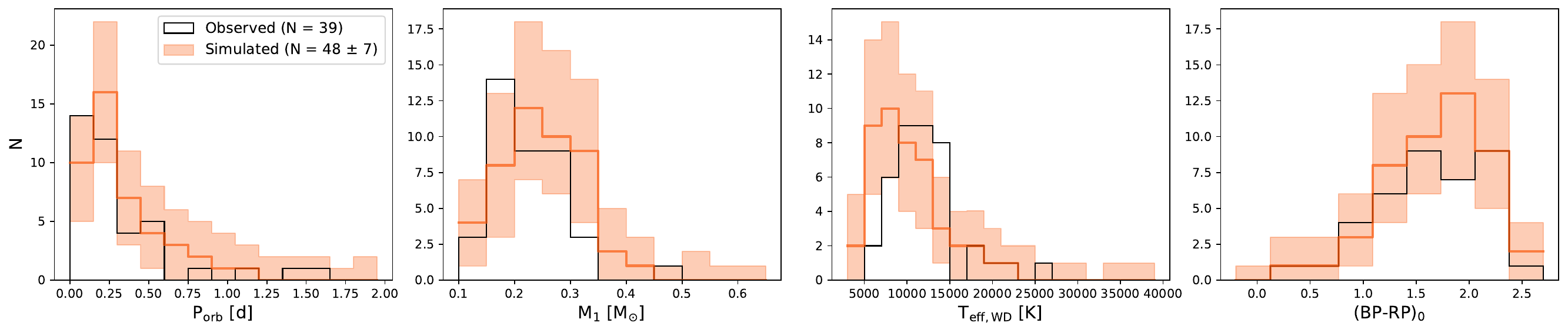}
    \caption{$P_{\rm orb}$, $M_1$, $T_{\rm eff, WD}$, and $(BP-RP)_0$ distributions of the observed (gray) and simulated (orange) EB samples. For each bin, we shade the region between the 5th and 95th percentiles covered by 1000 instances of the simulated sample generated via independent Bernoulli trials based on the detection probability (Figure \ref{fig:cmd_ebs}). The observed counts in most bins are consistent with the simulated counts, with notable exceptions at the drop-off for $M_1 \gtrsim 0.3\,M_{\odot}$ and $T_{\rm eff} \lesssim 10,000\,$K.}
    \label{fig:eb_results}
\end{figure*}

\begin{figure*}
    \centering
    \includegraphics[width=0.49\linewidth]{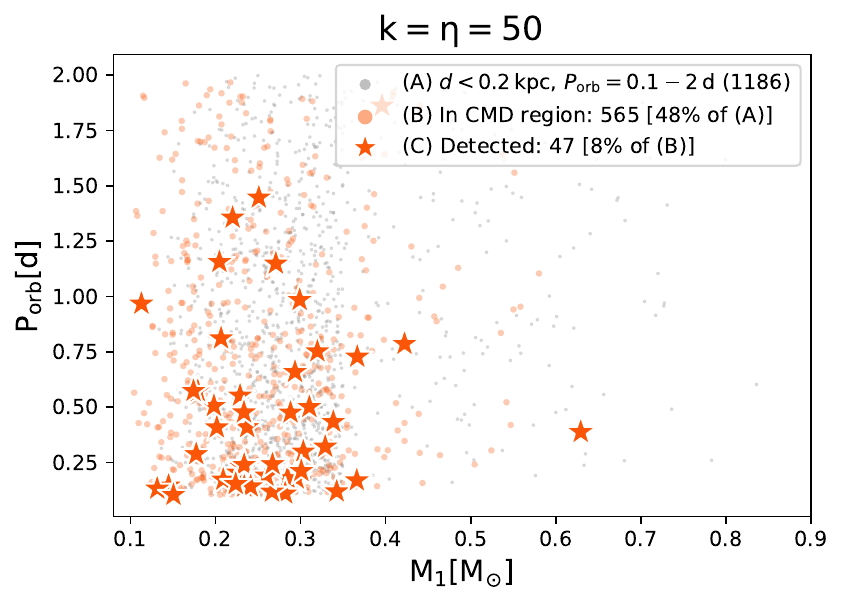}
    \includegraphics[width=0.49\linewidth]{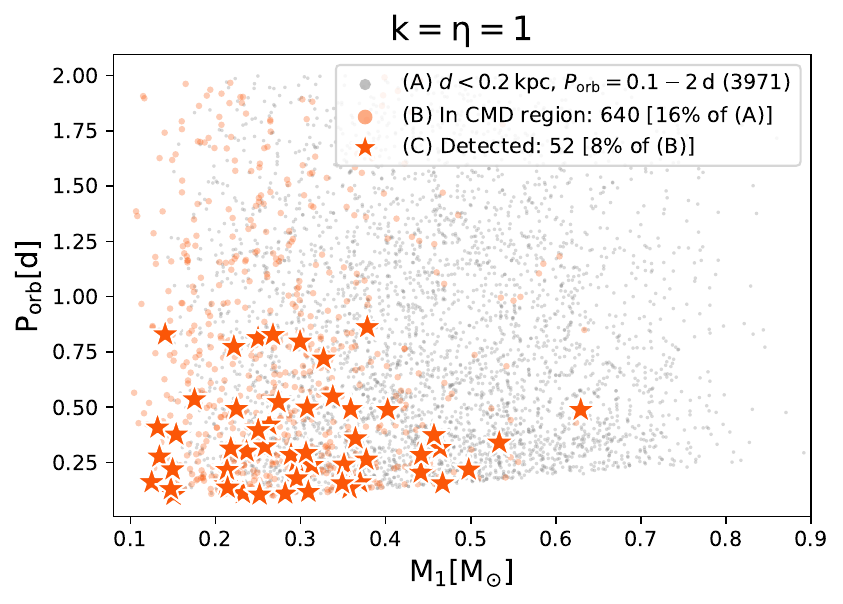}
    \caption{Simulated PCEBs on the $P_{\rm orb}-M_1$ space, assuming two MB prescriptions: $k=\eta=50$ (boosted and disrupted) and $k=\eta=1$ (neither boosted nor disrupted). The number of systems in the relevant CMD region (Figure \ref{fig:cmd_ebs}) and that are ultimately detected are similar between the two models. The difference is in the intrinsic number of all PCEBs. As expected, there are significantly more systems with primary masses above the fully convective limit of $0.35\,M_{\odot}$ in the case of no disruption, but these are disfavored by the CMD cut.}
    \label{fig:porb_m1_k_eta}
\end{figure*}

In Figure \ref{fig:eb_results}, we compare 1D distributions of orbital period, primary mass, WD temperature, and extinction-corrected color of the observed and simulated EB samples. Drawing independent Bernoulli trials, we generate 1000 realizations of the simulated sample using the detection probability plotted in Figure \ref{fig:cmd_ebs}. The shaded region shows the 5th–95th percentile range in each bin across all realizations. The expectation value and standard deviation of the total count ($48\pm7$) is also derived assuming Poisson-Bernoulli statistics. Therefore, the uncertainties quoted and plotted are only statistical errors and do not include model systematics. Still, the total number in the observed sample ($39$) is within $1.5\,\sigma$ of the predicted value. The simulated EB sample is made up of approximately equal numbers of post-RGB and post-AGB MT systems. 

The observed $P_{\rm orb}$ distribution is within the $90\%$ confidence interval of the simulated distribution across all bins. Although the same is true for $M_1$ given the uncertainties, there is possible evidence of a sharper decline at $\sim 0.28\,M_{\odot}$ in the observed sample than predicted by our model. Noting this, \citetalias{Shariat2026PASP} suggested that this may be attributed to MB. However, our MB prescription is both boosted and disrupted by a large factor ($k = \eta = 50$). While it is possible to improve the match by lowering the boundary below which MB is disrupted to $0.28\,M_{\odot}$, this is lower than the fully convective limit \citep[e.g.][]{Chabrier1997A&A, Baraffe1998A&A}. If it indeed exists, the origin of the steep drop-off in the observed $M_1$ distribution is thus uncertain, but if it is due to MB, this would require disruption of MB to occur at a mass somewhat lower than the fully convective limit (of which there have also been hints in hot subdwarf binaries; \citealt{Blomberg2024PASP}).

As for $T_{\rm eff, WD}$, we find a modest discrepancy in the location of the peak. The simulated systems host cooler WDs compared to the observed systems. On the other hand, the $(BP-RP)_0$ distributions are similar between the two samples. This may suggest that either the WD evolutionary models are predicting our simulated WDs to be too blue at their given temperatures, and/or the temperatures of the observed WDs, derived from just two UV photometric points, are overestimated.

% Discuss how things change with different k and eta
The expected number of systems in the final sample is relatively insensitive to the chosen values of $k$ and $\eta$ for the MB prescription. In Figure \ref{fig:porb_m1_k_eta}, we plot orbital period and primary mass for the simulated population with $k=\eta=50$ (i.e., highly boosted and disrupted, the fiducial choice) and $k=\eta=1$ (i.e., not boosted nor disrupted). We see that the total number only increases from $\sim 50$ to $\sim 60$, going from one to the other. What varies between them is the fraction of all short-period ($P_{\rm orb} = 0.1- 2\,$d) PCEBs that are located in the specified region of the CMD (Figure \ref{fig:cmd_ebs}), which decreases from $\sim 50\%$ with $k=\eta=50$ to $\sim 20\%$ with $k=\eta=1$. Meanwhile, we find that both $k$ and $\eta$ must be large ($\gtrsim50$) for there to be a clear drop-off in the $M_1$ distribution beyond the fully convective boundary. Additionally, we find that taking $k < \eta$ results in a shallower $P_{\rm orb}$ distribution compared to $k = \eta$, but given the uncertainties on the observed distribution, we can not confidently rule out such a model.

\subsection{Properties of the intrinsic post-MT population} \label{ssec:intrinsic}

\begin{figure*}
    \centering
    \includegraphics[width=1.0\linewidth]{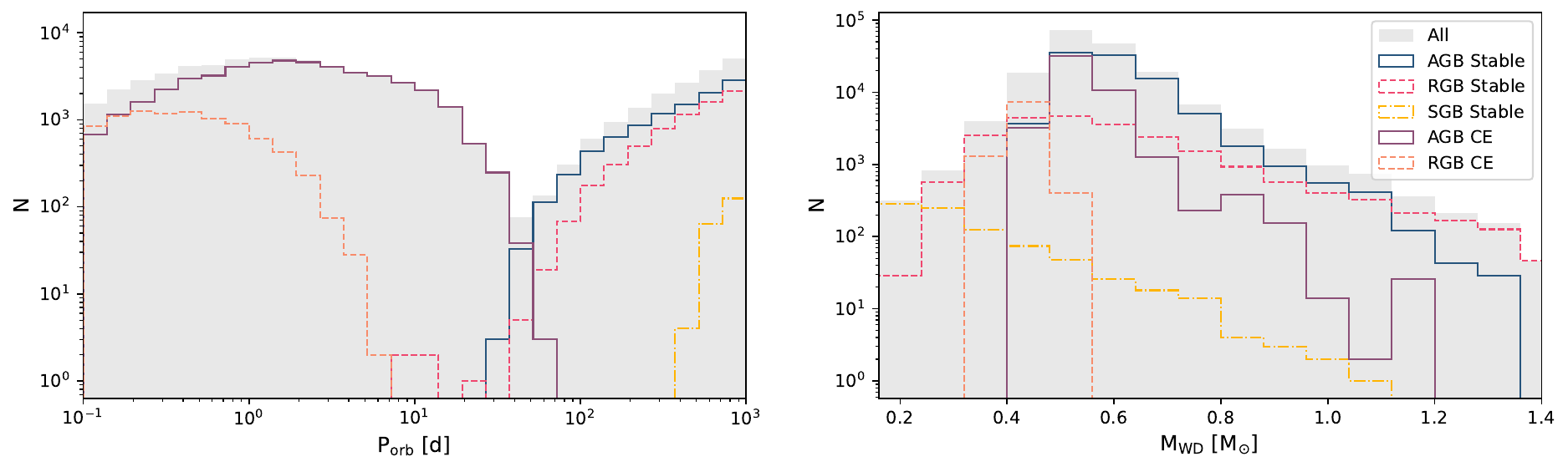}
    \caption{Distribution of orbital periods (\textit{Left}) and WD masses (\textit{Right}) of all post-interaction WDMS binaries within $500\,$pc according to our fiducial model. Stable MT products dominate at $P_{\rm orb} \sim 100\,d$, while PCEBs dominate at shorter periods. As our model considers truncation of the core growth at the onset of CEE, on average, PCEBs host lower mass WDs.}
    \label{fig:intrinsic}
\end{figure*}

\begin{figure*}
    \centering
    \includegraphics[width=1.0\linewidth]{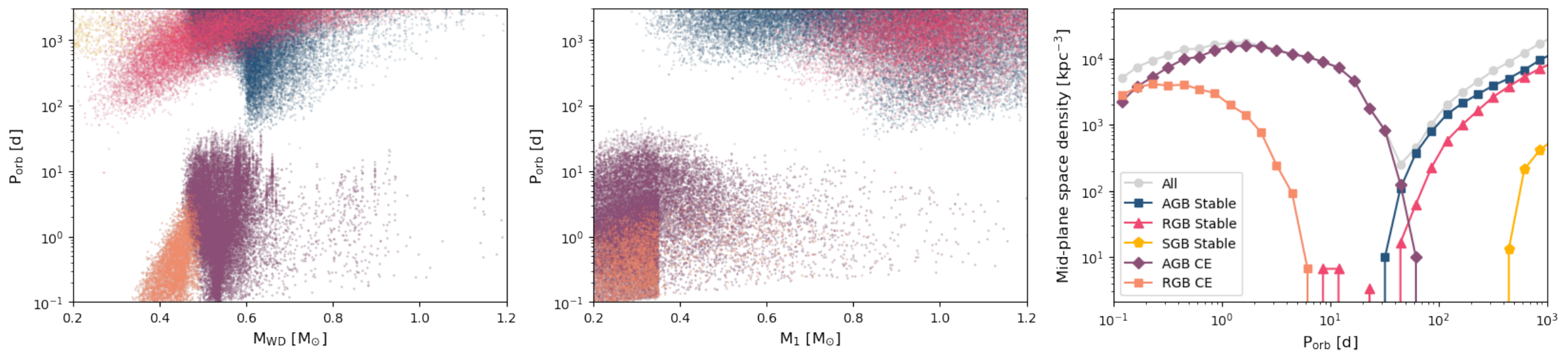}
    \caption{\textit{Left}: Orbital period against WD mass for the systems in Figure \ref{fig:intrinsic}. \textit{Center}: Orbital period against primary mass. The sharp boundary at $0.35\,M_{\odot}$ is the fully convective boundary below which MB is disrupted (Section \ref{sssec:MB}). \textit{Right}: Predicted space density of the same systems at the Galactic mid-plane as a function of orbital period. The predicted space density falls at periods intermediate to the astrometric and eclipsing populations, between $\sim 10$ and $100$\,d.}
    \label{fig:intrinsic_2}
\end{figure*}

In Figure \ref{fig:intrinsic}, we plot distributions of orbital period and WD mass of the full simulated post-MT WDMS binary population within $500\,$pc, showing contributions from products of the different MT processes. In the left panel of Figure \ref{fig:intrinsic_2}, we plot the same systems in the planes of orbital period and component masses. In the rightmost panel, we plot the inferred mid-plane space densities as a function of orbital period. Here, we include all WDMS binaries, including those with outer tertiaries. 

As expected, products of stable MT transfer are predicted to remain in wide orbits compared to PCEBs. Systems that interacted stably with SGB donors have the longest orbital periods as they host the lowest mass secondaries. We note that MT from SGB donors is predicted by \citet{Temmink2023AA} to be very stable, so no PCEBs are formed in this way. 

Overall, the population is dominated by products of MT from AGB donors across the range of orbital periods considered. Although the AGB phase is very short-lived, a star is also maximally radially extended during this phase, highlighting the importance of improving our understanding and treatment of MT from AGB donors. On the other hand, except at the lowest WD masses, there is minimal contribution from systems with SGB donors. 

Because the model truncates core growth at the onset of CEE, it predicts post-RGB common envelope products to have $M_{\rm WD} < 0.5\,M_{\odot}$ and most post-AGB common envelope products to have $M_{\rm WD}$ between $0.5$ and $0.7\,M_{\odot}$.

Lastly, the rate of PCEBs in our simulated population does not fall off sharply until $P_{\rm orb} \gtrsim 20\,$d. This is roughly consistent with the findings of \citet{Ashley2019MNRAS} whose search for PCEBs leads to an approximately log-uniform $P_{\rm orb}$ distribution up to $\sim 10\,$d, extending beyond the $2\,$d limit of \citet{Shariat2026arXiv}, although their work does not correct for the observational biases. 

\subsubsection{Intermediate orbital separations: Between PCEBs and stable MT products}

The model predicts a bimodal period distribution, with few WDMS binaries at intermediate periods of $\sim 10-100\,$d. Indeed, very few systems have so far been discovered at these periods. However, one of these systems, G203-47, stands out. G203-47 hosts a $\gtrsim 0.5\,M_{\odot}$ WD in a $14.9\,$d orbit around an M dwarf companion \citep{Delfosse1999A&A}. Located just $7.5\,$pc away, it is one of the ten nearest known WDs, suggesting that systems like it may be common. If we assume that it is the only one of its kind within $10\,$pc, then its inferred space density is $\sim 2\times10^{-4}\,\mathrm{pc}^{-3}$, which is about an order of magnitude higher than that from our simulated population. However, as pointed out by \citetalias{Shariat2026PASP}, this could be a one-off discovery, as a search for PCEBs conducted by \citet{Ashley2019MNRAS} did not find any systems with periods longer than $10\,$d, despite being sensitive to them. Recently, \citet{Lin2026ApJ} identified a second system, hosting a $\gtrsim 0.58\,M_{\odot}$ WD in a $14\,$d orbit around a K-type companion located $115\,$pc away. Additionally, several others in similar periods but hosting ultra-massive WD candidates ($\gtrsim 1.2\,M_{\odot}$) have been identified \citep{Wonnacott1993MNRAS, Yamaguchi2024MNRAS}. An extensive, volume-complete search targeting WDMS binaries with $P_{\rm orb}\sim 10-100\,$d would be beneficial in the near future to place stronger constraints on their statistics. Regarding their formation histories, several works suggest they are products of a common-envelope phase from a thermally pulsating AGB donor \citep[e.g.][]{Yamaguchi2024MNRAS, Belloni2024A&A, Yamaguchi2024PASP}. During this phase, the donor has an extremely loosely bound envelope, which can lead to very efficient envelope ejection and thus minimal orbital shrinkage. In this work, we assumed a single value of $\alpha\lambda = 0.3$. However, if the aforementioned interpretation is correct, larger values of $\lambda$ for more evolved AGB donors at the onset of MT could explain a higher occurrence rate of long-period PCEBs. Along the same lines, \citet{Torres2025A&A} argued that a single universal value of $\alpha$ is insufficient to explain their sample of observed systems \citep{Brown2023MNRAS}, instead requiring a range of values and finding a mild anti-correlation between $\alpha$ and the MS star mass. We caution that, in our model, we did not independently vary $\alpha$ or $\lambda$, as they are fully degenerate. However, the two parameters represent physically distinct quantities as described in Section \ref{sssec:cee}.

\section{Conclusion} \label{sec:conclusions}

In this work, we forward modeled three observed samples of post-MT WDMS binaries with well-characterized selection functions: (1) the ``AMRF" sample, constructed almost solely based on astrometric orbits from Gaia DR3 \citep{Shahaf2024MNRAS} and dominated by stable MT products from AGB donors, (2) the ``UV excess" sample, where hot WDs were identified based on the GALEX FUV excess and which is primarily made up of post-RGB stable MT products (Section \ref{ssec:uv_observed}), and lastly, (3) the ``EB" sample from \citetalias{Shariat2026PASP}, consisting of short-period PCEBs selected based on their Gaia CMD positions and eclipses in ZTF light curves (Section \ref{ssec:eb_observed}). Together, they probe a wide range of MT histories. Starting from a simulated population of zero-age binaries, we employed simple analytic models of binary evolution to get to the present-day population. Selection cuts were then applied to these present-day systems to generate simulated samples whose properties could then be compared to the true datasets. By simultaneously modeling multiple datasets, we attempt to construct a unified model of binary MT that probes a wide range of MT processes. We summarize our main findings below: 
\begin{itemize}
    \item To reproduce both the total number and WD mass distributions of the observed samples, we find that MT from AGB donors must be more stable than that from RGB donors in terms of the critical mass ratios. For AGB donors, we implement a fixed (accretor-to-donor) critical mass ratio of $0.4$. Meanwhile, for SGB and RGB donors, we implement quasi-adiabatic critical mass ratios derived by \citet{Temmink2023AA}. For RGB donors, we implement a lower bound of $0.65$ which is necessary to reduce contribution from low-mass WDs in the AMRF sample. (Section \ref{sssec:stability}) 
    \item For systems that experience stable MT, we evolve orbits according to relations from \citet{Soberman1997A&A} assuming fully non-conservative MT via isotropic winds from the accretor (Section \ref{sssec:stable_mt_agb}, \ref{sssec:stable_mt_rgb}). This leads to good agreement between the simulated and observed orbital period distributions of both post-AGB and post-RGB samples (Section \ref{sssec:comparison_amrf}, \ref{sssec:comparison_uv_excess}). 
    \item Products of stable MT from RGB and AGB donors are assumed to retain some of their initial eccentricities (Section \ref{sssec:stable_mt_agb}, \ref{sssec:stable_mt_rgb}). This is required to reproduce the non-zero peaks in the observed eccentricities of systems in both populations, suggesting inefficient tidal circularization and/or eccentricity pumping mechanisms taking place. In particular, the eccentricities of UV excess binaries hosting low-mass WDs are in contrast to the population of millisecond pulsar + He WD binaries in near-circular orbits, warranting a more in-depth look into the source of their large eccentricities in the near future (Section \ref{sssec:comparison_amrf},  \ref{sssec:comparison_uv_excess}). 
    \item We over-predict the number of post-RGB stable MT systems hosting primaries $\gtrsim 1.4\,M_{\odot}$ nearing the end of their main sequence lifetime that enter the UV excess sample. The reason for the lack of such systems in the observed sample remains unclear. In the near future, UV spectra will be obtained for some of the observed systems to confirm the nature of the hot secondary and optical radial velocity follow-up will allow us to validate the Gaia orbital solutions (Section \ref{sssec:comparison_uv_excess}).  
    \item For systems that experience unstable MT and enter a common envelope phase, we evolve orbits according to the $\alpha\lambda$-formalism with a fixed efficiency of $\alpha\lambda=0.3$ (Section \ref{sssec:cee}). The resulting PCEBs with short post-MT periods also experience further orbital shrinkage via MB. For this, we implement the prescription for saturated MB introduced by \citet{Belloni2024A&A} that can be both boosted in strength and disrupted below the fully convective boundary (Section \ref{sssec:MB}). We find that this allows us to reproduce both the number and orbital period distribution of the EB sample (Section \ref{sssec:comparison_eb}). 
    \item We find weak evidence of a discrepancy between the mass distribution of M dwarfs in the observed and simulated EB sample. (Section \ref{sssec:comparison_eb}; Figure \ref{fig:eb_results}). If this feature does in fact exist, its origin is unclear. While the disrupted MB prescription does lead to the sample being dominated by systems with primary masses below the fully convective boundary of $\sim 0.35\,M_{\odot}$, this will not be sufficient to reproduce the very steep cutoff in the observed distribution above a lower mass of $\sim 0.28\,M_{\odot}$. 
    \item Our fiducial model predicts that the population of post-MT WDMS binaries with $P_{\rm orb}=100-1000\,$d is dominated by MT products from AGB donors (Figure \ref{fig:intrinsic}). We also constrain space densities as a function of orbital period (Figure \ref{fig:intrinsic_2}) and predict an intrinsic dearth of systems at intermediate period ranges of $\sim 10-100\,$d (Section \ref{ssec:intrinsic}). However, the rate of such systems could be higher if more efficient common envelope ejection is assumed for a subset of systems which end up ``wide" PCEBs. Future searches targeting WDMS binaries at these orbital periods will help in placing stronger constraints.
\end{itemize}

\section{Acknowledgments}

This research was supported by NSF grants AST-2307232 and AST-2540180. NY acknowledges support from the Ezoe Memorial Recruit Foundation scholarship.
CS acknowledges support from the Department of Energy Computational Science Graduate Fellowship, supported by the U.S. Department of Energy, Office of Science, Office of Advanced Scientific Computing Research, under Award Number DE-SC0026073. 

This work has made use of data from the European Space Agency (ESA) mission {\it Gaia} (\url{https://www.cosmos.esa.int/gaia}), processed by the {\it Gaia} Data Processing and Analysis Consortium (DPAC,
\url{https://www.cosmos.esa.int/web/gaia/dpac/consortium}). Funding for the DPAC has been provided by national institutions, in particular the institutions participating in the {\it Gaia} Multilateral Agreement.

\vspace{5mm}
\facilities{Gaia}

\software{}

\appendix

\section{Different models of post-MT eccentricities} \label{appendix:ecc}

\begin{figure*}
    \centering
    \includegraphics[width=1.0\linewidth]{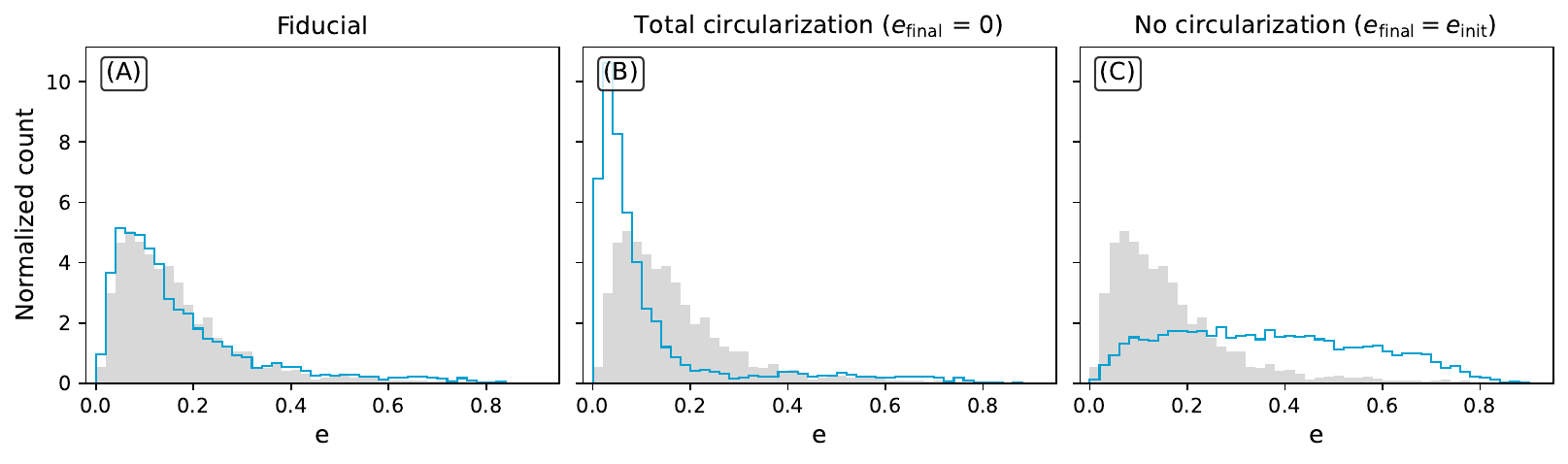}
    \caption{Comparison of the predicted eccentricity distribution for the simulated astrometric samples assuming different levels of circularization. Panel A is our fiducial model, where initial eccentricity is partially retained post-MT. In panel B is the result assuming complete circularization (i.e. zero final eccentricities; in this case, the non-zero measured eccentricities are a result of measurement errors). Panel C assumes no change in the eccentricities from initial values (i.e. no circularization). For comparison, the distribution of the true observed sample is plotted in gray. All histograms have been scaled to the same normalization to isolate differences in their shapes. }
    \label{fig:ecc_models}
\end{figure*}

In Figure \ref{fig:ecc_models}, we show plots comparing the eccentricity distributions of the observed and simulated NCE samples, assuming (1) total circularization and (2) eccentricities unchanged from initial values (i.e. no circularization). We also compare distributions to those resulting from fiducial model, where we implement an empirical relation between the pre- and post-MT eccentricities (Section \ref{sssec:stable_mt_agb}) that results in an intrinsic post-MT eccentricity distribution strongly peaked at $\sim 0.2$. 

As eccentricity cannot be negative, zero true eccentricities are pushed towards positive measured values. Still, complete circularization results in a very sharp peak close to zero, which is discrepant to the observed sample. Meanwhile, we see that assuming no change in eccentricity from their initial value significantly over-predicts the fraction of systems with $e\gtrsim 0.3$. These results reflect the fact that Gaia's sensitivity to binary orbits is only weakly sensitive to eccentricity at moderate eccentricities (e.g. \citealt{Wu2025ApJL}).

\section{Color excess calculation} \label{appendix:ce_calc}

\begin{figure*}
    \centering
    \includegraphics[width=1.0\linewidth]{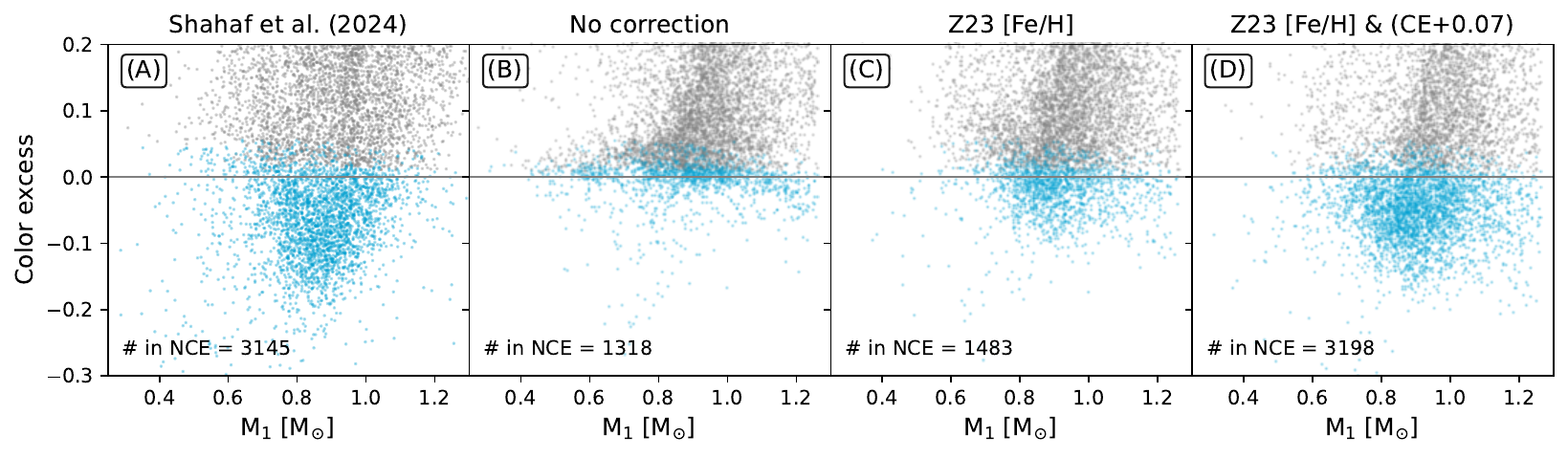}
    \caption{CE against primary mass for systems in the observed (\citealt{Shahaf2024MNRAS}; panel A) and simulated  (panels B - D) NCI (gray + blue) and NCE samples (blue). The observed points form a U-shaped trend towards negative color excesses for $M_1 \sim 0.7-1.0\,M_{\odot}$. In panel B is the result for the simulated population without any correction to the ``measured" [Fe/H] values. As expected, there is a concentration of systems with color excess$\sim0$ corresponding to the WD+MS binaries and few systems with color excess$\lesssim -0.05$, in contrast to the observed sample. In panel C, we have applied an offset to mimic the systematic error in the [Fe/H] values derived by \citet{Zhang2023MNRAS} (Figure \ref{fig:z23_offset}). This forms the U-shaped trend at the same range of primary masses as the observed sample, but with an overall offset to more positive color excesses. In panel D, we additionally apply a constant correction of $-0.07\,$dex to the calculated color excesses.}
    \label{fig:ce_corr}
\end{figure*}

\begin{figure*}
    \centering
    \includegraphics[width=0.65\linewidth]{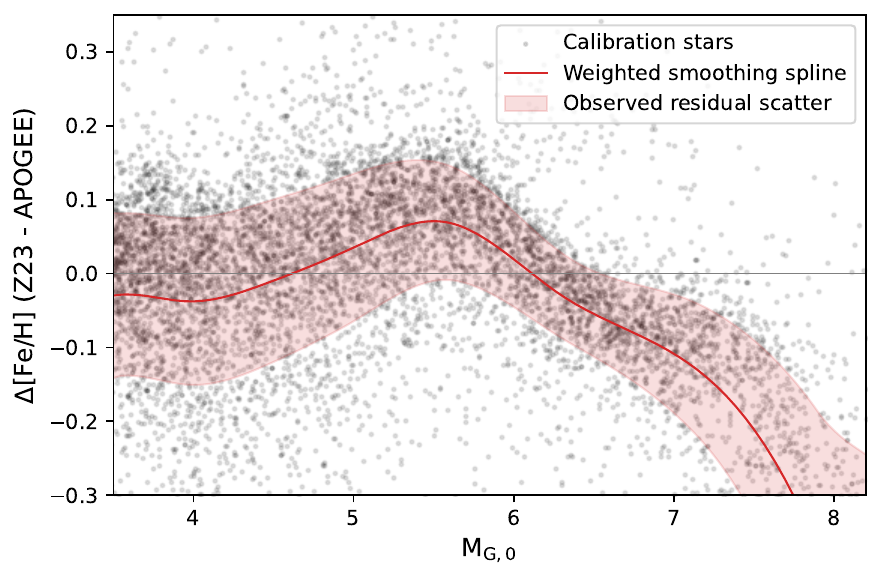}
    \caption{Difference in the [Fe/H] derived by \citet{Zhang2023MNRAS} and APOGEE for the same sources plotted against their absolute, extinction-corrected, G-band magnitude. The [Fe/H] from \citet{Zhang2023MNRAS} are more positive by up to $\sim 0.1\,$dex for $M_{\rm G, 0} \sim 4.5-6$, precisely corresponding to the range of MS star masses where the negative color excesses are found for the observed sample (Panel A; Figure \ref{fig:ce_corr}). The points are fitted with a weighted smoothing spline which is used to correct the measured [Fe/H] of the simulated NCI sample in calculating the expected color. Residuals from the best-fit spline is modeled as a Gaussian.}
    \label{fig:z23_offset}
\end{figure*}

Here, we detail the changes we made to the calculation of the color excess from the work of \citet{Yamaguchi2025PASP} (Section \ref{ssec:model_sel_amrf})

In panel A of Figure \ref{fig:ce_corr}, we plot the color excess against primary mass for the true NCI sample \citep{Shahaf2024MNRAS}. Those with positive values are expected to be triples where the secondary, made up of two low-mass MS stars, contributes little to the total flux in comparison to the primary MS star but significantly reddens their combined light. On the other hand, the source of the systematically negative color excess values for systems with $M_1\sim0.7-1.0\,M_{\odot}$ is not immediately obvious. Systems hosting compact objects which contribute little to no light in the optical are naively expected to have zero color excess. Indeed, this is what we get from in the simulated NCI sample without any correction (panel B; Figure \ref{fig:ce_corr}). Thus, a negative color excess indicates the presence of an additional component that is fainter but significantly bluer than the MS primary. We note the while hot WDs can contribute to the blue optical, effective temperatures of $\gtrsim 50,000\,$K are necessary to explain the observed offset and few such WDs should be present. Therefore, we conclude that this issue likely stems from the color excess calculation. 

This problem was previously noted by \citet{Yamaguchi2025PASP}, who suggested that the cause may be in the predicted MS evolution of the $(B-I)$ color by the PARSEC models and who applied a purely empirical correction on the color excess as function of primary mass to reproduce the observed trend. In this section, we demonstrate that the discrepancy instead originates primarily from a systematic error in the [Fe/H] values of \citet{Zhang2023MNRAS} used by \citet{Shahaf2024MNRAS} to calculate expected colors. 

This is illustrated in Figure \ref{fig:z23_offset}, where we plot the difference in [Fe/H] derived by \citet{Zhang2023MNRAS} and APOGEE \citep{Majewski2017AJ, Nidever2015AJ, Wilson2019PASP, GarciaPerez2016AJ, Smith2021AJ, Shetrone2015ApJS}, $\Delta\mathrm{[Fe/H]}$ for the same stars, as a function of the absolute extinction-corrected G-band magnitude, $M_{\rm G, 0}$. We cross-matched the two catalogs, selecting a random subset of $\sim 10,000$ stars based on their Gaia DR3 source ID. There is an offset to positive values for $M_{\rm G, 0} \sim 4.5-6$ which precisely corresponds to the range of MS star mass in the observed NCI sample with a trend toward negative color excesses. In theory, for a solar-type star of fixed mass on the MS, its color is expected to become redder at higher [Fe/H]. Therefore, an overestimated [Fe/H] would lead to an expected color that is too red and consequently, to negative color excess. We find qualitatively similar systematic trends in $\Delta\mathrm{[Fe/H]}$ calculating via cross-match to the GALAH DR3 survey \citep{DeSilva2015MNRAS, Buder2021MNRAS} as well as to the catalog from \citet{Andrae2023ApJS}. 

We fit the offset with APOGEE plotted in Figure \ref{fig:z23_offset} using a weighted smoothing spline, and model the residuals as Gaussian noise. We map this fit to the simulated [Fe/H] values to mimic \citet{Zhang2023MNRAS} which are used in calculating the expected $(B-I)$ colors. As done by \citet{Yamaguchi2025PASP}, we fix the error on these ``measured" [Fe/H] to $0.04\,$dex, and assign [Fe/H]$=0$ to a random $20\%$ of the population with an enhanced error of $0.25\,$dex. The latter follows the treatment taken by \citet{Shahaf2024MNRAS} of $\sim 20\%$ of their sources which are flagged as having unreliable parameters by \citealt{Zhang2023MNRAS} (i.e. \texttt{quality\_flag}$\geq8$). 

In panel C of Figure \ref{fig:ce_corr}, we plot results after this ``correction" has been made. This introduces the U-shaped group of negative color excesses for primary masses $\sim 0.7-1.0\,M_{\odot}$, similar to the observed NCI sample. However, the location of the minimum is offset by a fixed value of $\sim 0.07$ in color excess, which we additionally correct for in panel D. The constant offset of $< 0.1\,$mag in color can be reasonably attributed to an error in the model predictions of the absolute color. In fact, PARSEC and MIST evolutionary grids generally predict very close agreement in the predicted evolution of $(B-I)$ with age on the MS, but often with a roughly constant discrepancy in $(B-I)$ at any given age on the order of $\sim 0.1\,$dex. 

While this change in the color excess calculation minimally affects the inferred parameter distributions, the fraction of triples that enter the final NCE sample is now larger, making up $\sim7\%$ of the total number compared to $\sim0.5\%$ in \citet{Yamaguchi2025PASP}. This suggests that a possibly stricter cut for the observed sample may be warranted, though the amount of contamination is sensitive to the chosen value of the fixed color excess offset.

\bibliographystyle{aasjournal}

\begin{thebibliography}{}
\expandafter\ifx\csname natexlab\endcsname\relax\def\natexlab#1{#1}\fi
\providecommand{\url}[1]{\href{#1}{#1}}
\providecommand{\dodoi}[1]{doi:~\href{http://doi.org/#1}{\nolinkurl{#1}}}
\providecommand{\doeprint}[1]{\href{http://ascl.net/#1}{\nolinkurl{http://ascl.net/#1}}}
\providecommand{\doarXiv}[1]{\href{https://arxiv.org/abs/#1}{\nolinkurl{https://arxiv.org/abs/#1}}}

\bibitem[{{Abbott} {et~al.}(2016){Abbott}, {Abbott}, {Abbott}, {Abernathy}, {Acernese}, {Ackley}, {Adams}, {Adams}, {Addesso}, {Adhikari}, {Adya}, {Affeldt}, {Agathos}, {Agatsuma}, {Aggarwal}, {Aguiar}, {Aiello}, {Ain}, {Ajith}, {Allen}, {Allocca}, {Altin}, {Anderson}, {Anderson}, {Arai}, {Arain}, {Araya}, {Arceneaux}, {Areeda}, {Arnaud}, {Arun}, {Ascenzi}, {Ashton}, {Ast}, {Aston}, {Astone}, {Aufmuth}, {Aulbert}, {Babak}, {Bacon}, {Bader}, {Baker}, {Baldaccini}, {Ballardin}, {Ballmer}, {Barayoga}, {Barclay}, {Barish}, {Barker}, {Barone}, {Barr}, {Barsotti}, {Barsuglia}, {Barta}, {Bartlett}, {Barton}, {Bartos}, {Bassiri}, {Basti}, {Batch}, {Baune}, {Bavigadda}, {Bazzan}, {Behnke}, {Bejger}, {Belczynski}, {Bell}, {Bell}, {Berger}, {Bergman}, {Bergmann}, {Berry}, {Bersanetti}, {Bertolini}, {Betzwieser}, {Bhagwat}, {Bhandare}, {Bilenko}, {Billingsley}, {Birch}, {Birney}, {Birnholtz}, {Biscans}, {Bisht}, {Bitossi}, {Biwer}, {Bizouard}, {Blackburn}, {Blair}, {Blair}, {Blair}, {Bloemen}, {Bock}, {Bodiya}, {Boer},
  {Bogaert}, {Bogan}, {Bohe}, {Bojtos}, {Bond}, {Bondu}, {Bonnand}, {Boom}, {Bork}, {Boschi}, {Bose}, {Bouffanais}, {Bozzi}, {Bradaschia}, {Brady}, {Braginsky}, {Branchesi}, {Brau}, {Briant}, {Brillet}, {Brinkmann}, {Brisson}, {Brockill}, {Brooks}, {Brown}, {Brown}, {Brown}, {Buchanan}, {Buikema}, {Bulik}, {Bulten}, {Buonanno}, {Buskulic}, {Buy}, {Byer}, {Cabero}, {Cadonati}, {Cagnoli}, {Cahillane}, {Bustillo}, {Callister}, {Calloni}, {Camp}, {Cannon}, {Cao}, {Capano}, {Capocasa}, {Carbognani}, {Caride}, {Diaz}, {Casentini}, {Caudill}, {Cavagli{\`a}}, {Cavalier}, {Cavalieri}, {Cella}, {Cepeda}, {Baiardi}, {Cerretani}, {Cesarini}, {Chakraborty}, {Chalermsongsak}, {Chamberlin}, {Chan}, {Chao}, {Charlton}, {Chassande-Mottin}, {Chen}, {Chen}, {Cheng}, {Chincarini}, {Chiummo}, {Cho}, {Cho}, {Chow}, {Christensen}, {Chu}, {Chua}, {Chung}, {Ciani}, {Clara}, {Clark}, {Cleva}, {Coccia}, {Cohadon}, {Colla}, {Collette}, {Cominsky}, {Constancio}, {Conte}, {Conti}, {Cook}, {Corbitt}, {Cornish}, {Corsi}, {Cortese}, {Costa},
  {Coughlin}, {Coughlin}, {Coulon}, {Countryman}, {Couvares}, {Cowan}, {Coward}, \& {Cowart}}]{Abbott2016PhRvL}
{Abbott}, B.~P., {Abbott}, R., {Abbott}, T.~D., {et~al.} 2016, \prl, 116, 061102, \dodoi{10.1103/PhysRevLett.116.061102}

\bibitem[{{Alpar} {et~al.}(1982){Alpar}, {Cheng}, {Ruderman}, \& {Shaham}}]{Alpar1982Natur}
{Alpar}, M.~A., {Cheng}, A.~F., {Ruderman}, M.~A., \& {Shaham}, J. 1982, \nat, 300, 728, \dodoi{10.1038/300728a0}

\bibitem[{{Andrae} {et~al.}(2023){Andrae}, {Rix}, \& {Chandra}}]{Andrae2023ApJS}
{Andrae}, R., {Rix}, H.-W., \& {Chandra}, V. 2023, \apjs, 267, 8, \dodoi{10.3847/1538-4365/acd53e}

\bibitem[{{Arseneau} {et~al.}(2026){Arseneau}, {Hermes}, {Camisassa}, {Raddi}, \& {Bauer}}]{Arseneau2026arXiv}
{Arseneau}, S.~M., {Hermes}, J.~J., {Camisassa}, M.~E., {Raddi}, R., \& {Bauer}, E.~B. 2026, \apj, 1000, 297, \dodoi{10.3847/1538-4357/ae4c4b}

\bibitem[{{Ashley} {et~al.}(2019){Ashley}, {Farihi}, {Marsh}, {Wilson}, \& {G{\"a}nsicke}}]{Ashley2019MNRAS}
{Ashley}, R.~P., {Farihi}, J., {Marsh}, T.~R., {Wilson}, D.~J., \& {G{\"a}nsicke}, B.~T. 2019, \mnras, 484, 5362, \dodoi{10.1093/mnras/stz298}

\bibitem[{{Baraffe} {et~al.}(1998){Baraffe}, {Chabrier}, {Allard}, \& {Hauschildt}}]{Baraffe1998A&A}
{Baraffe}, I., {Chabrier}, G., {Allard}, F., \& {Hauschildt}, P.~H. 1998, \aap, 337, 403, \dodoi{10.48550/arXiv.astro-ph/9805009}

\bibitem[{{Bauer} {et~al.}(2026){Bauer}, {Dotter}, {Conroy}, {Cunningham}, {Park}, \& {Tremblay}}]{Bauer2026ApJS}
{Bauer}, E.~B., {Dotter}, A., {Conroy}, C., {et~al.} 2026, \apjs, 283, 41, \dodoi{10.3847/1538-4365/ae401e}

\bibitem[{{Bayo} {et~al.}(2008){Bayo}, {Rodrigo}, {Barrado Y Navascu{\'e}s}, {Solano}, {Guti{\'e}rrez}, {Morales-Calder{\'o}n}, \& {Allard}}]{Bayo2008A&A}
{Bayo}, A., {Rodrigo}, C., {Barrado Y Navascu{\'e}s}, D., {et~al.} 2008, \aap, 492, 277, \dodoi{10.1051/0004-6361:200810395}

\bibitem[{{B{\'e}dard} {et~al.}(2020){B{\'e}dard}, {Bergeron}, {Brassard}, \& {Fontaine}}]{Bergeron2020}
{B{\'e}dard}, A., {Bergeron}, P., {Brassard}, P., \& {Fontaine}, G. 2020, \apj, 901, 93, \dodoi{10.3847/1538-4357/abafbe}

\bibitem[{{Bellm} {et~al.}(2019){Bellm}, {Kulkarni}, {Graham}, {Dekany}, {Smith}, {Riddle}, {Masci}, {Helou}, {Prince}, {Adams}, {Barbarino}, {Barlow}, {Bauer}, {Beck}, {Belicki}, {Biswas}, {Blagorodnova}, {Bodewits}, {Bolin}, {Brinnel}, {Brooke}, {Bue}, {Bulla}, {Burruss}, {Cenko}, {Chang}, {Connolly}, {Coughlin}, {Cromer}, {Cunningham}, {De}, {Delacroix}, {Desai}, {Duev}, {Eadie}, {Farnham}, {Feeney}, {Feindt}, {Flynn}, {Franckowiak}, {Frederick}, {Fremling}, {Gal-Yam}, {Gezari}, {Giomi}, {Goldstein}, {Golkhou}, {Goobar}, {Groom}, {Hacopians}, {Hale}, {Henning}, {Ho}, {Hover}, {Howell}, {Hung}, {Huppenkothen}, {Imel}, {Ip}, {Ivezi{\'c}}, {Jackson}, {Jones}, {Juric}, {Kasliwal}, {Kaspi}, {Kaye}, {Kelley}, {Kowalski}, {Kramer}, {Kupfer}, {Landry}, {Laher}, {Lee}, {Lin}, {Lin}, {Lunnan}, {Giomi}, {Mahabal}, {Mao}, {Miller}, {Monkewitz}, {Murphy}, {Ngeow}, {Nordin}, {Nugent}, {Ofek}, {Patterson}, {Penprase}, {Porter}, {Rauch}, {Rebbapragada}, {Reiley}, {Rigault}, {Rodriguez}, {van Roestel}, {Rusholme}, {van
  Santen}, {Schulze}, {Shupe}, {Singer}, {Soumagnac}, {Stein}, {Surace}, {Sollerman}, {Szkody}, {Taddia}, {Terek}, {Van Sistine}, {van Velzen}, {Vestrand}, {Walters}, {Ward}, {Ye}, {Yu}, {Yan}, \& {Zolkower}}]{Bellm19}
{Bellm}, E.~C., {Kulkarni}, S.~R., {Graham}, M.~J., {et~al.} 2019, \pasp, 131, 018002, \dodoi{10.1088/1538-3873/aaecbe}

\bibitem[{{Belloni} {et~al.}(2024){Belloni}, {Schreiber}, {Moe}, {El-Badry}, \& {Shen}}]{Belloni2024A&A}
{Belloni}, D., {Schreiber}, M.~R., {Moe}, M., {El-Badry}, K., \& {Shen}, K.~J. 2024, \aap, 682, A33, \dodoi{10.1051/0004-6361/202347931}

\bibitem[{{Bhattacharjee} {et~al.}(2026){Bhattacharjee}, {El-Badry}, {Fuller}, {Shariat}, \& {Yamaguchi}}]{Bhattacharjee26}
{Bhattacharjee}, S., {El-Badry}, K., {Fuller}, J., {Shariat}, C., \& {Yamaguchi}, N. 2026, \pasp, 138, 084202, \dodoi{10.1088/1538-3873/ae8dff}

\bibitem[{{Bhattacharya} \& {van den Heuvel}(1991)}]{Bhattacharya1991PhR}
{Bhattacharya}, D., \& {van den Heuvel}, E.~P.~J. 1991, \physrep, 203, 1, \dodoi{10.1016/0370-1573(91)90064-S}

\bibitem[{{Blomberg} {et~al.}(2024){Blomberg}, {El-Badry}, {Breivik}, {Caiazzo}, {Nagarajan}, {Rodriguez}, {van Roestel}, {Vanderbosch}, \& {Yamaguchi}}]{Blomberg2024PASP}
{Blomberg}, L., {El-Badry}, K., {Breivik}, K., {et~al.} 2024, \pasp, 136, 124201, \dodoi{10.1088/1538-3873/ad94a2}

\bibitem[{{Bovy} {et~al.}(2016){Bovy}, {Rix}, {Green}, {Schlafly}, \& {Finkbeiner}}]{Bovy2016ApJ}
{Bovy}, J., {Rix}, H.-W., {Green}, G.~M., {Schlafly}, E.~F., \& {Finkbeiner}, D.~P. 2016, \apj, 818, 130, \dodoi{10.3847/0004-637X/818/2/130}

\bibitem[{{Breivik} {et~al.}(2020){Breivik}, {Coughlin}, {Zevin}, {Rodriguez}, {Kremer}, {Ye}, {Andrews}, {Kurkowski}, {Digman}, {Larson}, \& {Rasio}}]{Breivik2020ApJ}
{Breivik}, K., {Coughlin}, S., {Zevin}, M., {et~al.} 2020, \apj, 898, 71, \dodoi{10.3847/1538-4357/ab9d85}

\bibitem[{{Brown} {et~al.}(2023){Brown}, {Parsons}, {van Roestel}, {Rebassa-Mansergas}, {Breedt}, {Dhillon}, {Dyer}, {Green}, {Kerry}, {Littlefair}, {Marsh}, {Munday}, {Pelisoli}, {Sahman}, \& {Wild}}]{Brown2023MNRAS}
{Brown}, A.~J., {Parsons}, S.~G., {van Roestel}, J., {et~al.} 2023, \mnras, 521, 1880, \dodoi{10.1093/mnras/stad612}

\bibitem[{{Buder} {et~al.}(2021){Buder}, {Sharma}, {Kos}, {Amarsi}, {Nordlander}, {Lind}, {Martell}, {Asplund}, {Bland-Hawthorn}, {Casey}, {de Silva}, {D'Orazi}, {Freeman}, {Hayden}, {Lewis}, {Lin}, {Schlesinger}, {Simpson}, {Stello}, {Zucker}, {Zwitter}, {Beeson}, {Buck}, {Casagrande}, {Clark}, {{\v{C}}otar}, {da Costa}, {de Grijs}, {Feuillet}, {Horner}, {Kafle}, {Khanna}, {Kobayashi}, {Liu}, {Montet}, {Nandakumar}, {Nataf}, {Ness}, {Spina}, {Tepper-Garc{\'\i}a}, {Ting}, {Traven}, {Vogrin{\v{c}}i{\v{c}}}, {Wittenmyer}, {Wyse}, {{\v{Z}}erjal}, \& {Galah Collaboration}}]{Buder2021MNRAS}
{Buder}, S., {Sharma}, S., {Kos}, J., {et~al.} 2021, \mnras, 506, 150, \dodoi{10.1093/mnras/stab1242}

\bibitem[{{Castelli} \& {Kurucz}(2003)}]{Castelli2003IAUS}
{Castelli}, F., \& {Kurucz}, R.~L. 2003, in IAU Symposium, Vol. 210, Modelling of Stellar Atmospheres, ed. N.~{Piskunov}, W.~W. {Weiss}, \& D.~F. {Gray}, A20, \dodoi{10.48550/arXiv.astro-ph/0405087}

\bibitem[{{Chaboyer} {et~al.}(1995){Chaboyer}, {Demarque}, \& {Pinsonneault}}]{Chaboyer1995ApJ}
{Chaboyer}, B., {Demarque}, P., \& {Pinsonneault}, M.~H. 1995, \apj, 441, 865, \dodoi{10.1086/175408}

\bibitem[{{Chabrier} \& {Baraffe}(1997)}]{Chabrier1997A&A}
{Chabrier}, G., \& {Baraffe}, I. 1997, \aap, 327, 1039, \dodoi{10.48550/arXiv.astro-ph/9704118}

\bibitem[{{Chen} {et~al.}(2015){Chen}, {Bressan}, {Girardi}, {Marigo}, {Kong}, \& {Lanza}}]{Chen2015MNRAS}
{Chen}, Y., {Bressan}, A., {Girardi}, L., {et~al.} 2015, \mnras, 452, 1068, \dodoi{10.1093/mnras/stv1281}

\bibitem[{{Chen} {et~al.}(2014){Chen}, {Girardi}, {Bressan}, {Marigo}, {Barbieri}, \& {Kong}}]{Chen2014MNRAS}
{Chen}, Y., {Girardi}, L., {Bressan}, A., {et~al.} 2014, \mnras, 444, 2525, \dodoi{10.1093/mnras/stu1605}

\bibitem[{{Choi} {et~al.}(2016){Choi}, {Dotter}, {Conroy}, {Cantiello}, {Paxton}, \& {Johnson}}]{Choi2016ApJ}
{Choi}, J., {Dotter}, A., {Conroy}, C., {et~al.} 2016, \apj, 823, 102, \dodoi{10.3847/0004-637X/823/2/102}

\bibitem[{{Corongiu} {et~al.}(2012){Corongiu}, {Burgay}, {Possenti}, {Camilo}, {D'Amico}, {Lyne}, {Manchester}, {Sarkissian}, {Bailes}, {Johnston}, {Kramer}, \& {van Straten}}]{Corongiu2012ApJ}
{Corongiu}, A., {Burgay}, M., {Possenti}, A., {et~al.} 2012, \apj, 760, 100, \dodoi{10.1088/0004-637X/760/2/100}

\bibitem[{{Davis} {et~al.}(2012){Davis}, {Kolb}, \& {Knigge}}]{Davis2012MNRAS}
{Davis}, P.~J., {Kolb}, U., \& {Knigge}, C. 2012, \mnras, 419, 287, \dodoi{10.1111/j.1365-2966.2011.19690.x}

\bibitem[{{de Kool}(1990)}]{deKool1990ApJ}
{de Kool}, M. 1990, \apj, 358, 189, \dodoi{10.1086/168974}

\bibitem[{{De Silva} {et~al.}(2015){De Silva}, {Freeman}, {Bland-Hawthorn}, {Martell}, {de Boer}, {Asplund}, {Keller}, {Sharma}, {Zucker}, {Zwitter}, {Anguiano}, {Bacigalupo}, {Bayliss}, {Beavis}, {Bergemann}, {Campbell}, {Cannon}, {Carollo}, {Casagrande}, {Casey}, {Da Costa}, {D'Orazi}, {Dotter}, {Duong}, {Heger}, {Ireland}, {Kafle}, {Kos}, {Lattanzio}, {Lewis}, {Lin}, {Lind}, {Munari}, {Nataf}, {O'Toole}, {Parker}, {Reid}, {Schlesinger}, {Sheinis}, {Simpson}, {Stello}, {Ting}, {Traven}, {Watson}, {Wittenmyer}, {Yong}, \& {{\v{Z}}erjal}}]{DeSilva2015MNRAS}
{De Silva}, G.~M., {Freeman}, K.~C., {Bland-Hawthorn}, J., {et~al.} 2015, \mnras, 449, 2604, \dodoi{10.1093/mnras/stv327}

\bibitem[{{Delfosse} {et~al.}(1999){Delfosse}, {Forveille}, {Beuzit}, {Udry}, {Mayor}, \& {Perrier}}]{Delfosse1999A&A}
{Delfosse}, X., {Forveille}, T., {Beuzit}, J.-L., {et~al.} 1999, \aap, 344, 897, \dodoi{10.48550/arXiv.astro-ph/9812008}

\bibitem[{{Drimmel} {et~al.}(2003){Drimmel}, {Cabrera-Lavers}, \& {L{\'o}pez-Corredoira}}]{Drimmel2003A&A}
{Drimmel}, R., {Cabrera-Lavers}, A., \& {L{\'o}pez-Corredoira}, M. 2003, \aap, 409, 205, \dodoi{10.1051/0004-6361:20031070}

\bibitem[{{El-Badry} {et~al.}(2022){El-Badry}, {Conroy}, {Fuller}, {Kiman}, {van Roestel}, {Rodriguez}, \& {Burdge}}]{El-Badry2022MNRAS}
{El-Badry}, K., {Conroy}, C., {Fuller}, J., {et~al.} 2022, \mnras, 517, 4916, \dodoi{10.1093/mnras/stac2945}

\bibitem[{{El-Badry} {et~al.}(2024){El-Badry}, {Lam}, {Holl}, {Halbwachs}, {Rix}, {Mazeh}, \& {Shahaf}}]{El-Badry2024OJAp}
{El-Badry}, K., {Lam}, C., {Holl}, B., {et~al.} 2024, The Open Journal of Astrophysics, 7, 100, \dodoi{10.33232/001c.125461}

\bibitem[{{Gaia Collaboration} {et~al.}(2023{\natexlab{a}}){Gaia Collaboration}, {Vallenari}, {Brown}, {Prusti}, {de Bruijne}, {Arenou}, {Babusiaux}, {Biermann}, {Creevey}, {Ducourant}, {Evans}, {Eyer}, {Guerra}, {Hutton}, {Jordi}, {Klioner}, {Lammers}, {Lindegren}, {Luri}, {Mignard}, {Panem}, {Pourbaix}, {Randich}, {Sartoretti}, {Soubiran}, {Tanga}, {Walton}, {Bailer-Jones}, {Bastian}, {Drimmel}, {Jansen}, {Katz}, {Lattanzi}, {van Leeuwen}, {Bakker}, {Cacciari}, {Casta{\~n}eda}, {De Angeli}, {Fabricius}, {Fouesneau}, {Fr{\'e}mat}, {Galluccio}, {Guerrier}, {Heiter}, {Masana}, {Messineo}, {Mowlavi}, {Nicolas}, {Nienartowicz}, {Pailler}, {Panuzzo}, {Riclet}, {Roux}, {Seabroke}, {Sordo}, {Th{\'e}venin}, {Gracia-Abril}, {Portell}, {Teyssier}, {Altmann}, {Andrae}, {Audard}, {Bellas-Velidis}, {Benson}, {Berthier}, {Blomme}, {Burgess}, {Busonero}, {Busso}, {C{\'a}novas}, {Carry}, {Cellino}, {Cheek}, {Clementini}, {Damerdji}, {Davidson}, {de Teodoro}, {Nu{\~n}ez Campos}, {Delchambre}, {Dell'Oro}, {Esquej},
  {Fern{\'a}ndez-Hern{\'a}ndez}, {Fraile}, {Garabato}, {Garc{\'\i}a-Lario}, {Gosset}, {Haigron}, {Halbwachs}, {Hambly}, {Harrison}, {Hern{\'a}ndez}, {Hestroffer}, {Hodgkin}, {Holl}, {Jan{\ss}en}, {Jevardat de Fombelle}, {Jordan}, {Krone-Martins}, {Lanzafame}, {L{\"o}ffler}, {Marchal}, {Marrese}, {Moitinho}, {Muinonen}, {Osborne}, {Pancino}, {Pauwels}, {Recio-Blanco}, {Reyl{\'e}}, {Riello}, {Rimoldini}, {Roegiers}, {Rybizki}, {Sarro}, {Siopis}, {Smith}, {Sozzetti}, {Utrilla}, {van Leeuwen}, {Abbas}, {{\'A}brah{\'a}m}, {Abreu Aramburu}, {Aerts}, {Aguado}, {Ajaj}, {Aldea-Montero}, {Altavilla}, {{\'A}lvarez}, {Alves}, {Anders}, {Anderson}, {Anglada Varela}, {Antoja}, {Baines}, {Baker}, {Balaguer-N{\'u}{\~n}ez}, {Balbinot}, {Balog}, {Barache}, {Barbato}, {Barros}, {Barstow}, {Bartolom{\'e}}, {Bassilana}, {Bauchet}, {Becciani}, {Bellazzini}, {Berihuete}, {Bernet}, {Bertone}, {Bianchi}, {Binnenfeld}, {Blanco-Cuaresma}, {Blazere}, {Boch}, {Bombrun}, {Bossini}, {Bouquillon}, {Bragaglia}, {Bramante}, {Breedt},
  {Bressan}, {Brouillet}, {Brugaletta}, {Bucciarelli}, {Burlacu}, {Butkevich}, {Buzzi}, {Caffau}, {Cancelliere}, {Cantat-Gaudin}, {Carballo}, {Carlucci}, {Carnerero}, {Carrasco}, {Casamiquela}, {Castellani}, {Castro-Ginard}, {Chaoul}, {Charlot}, {Chemin}, {Chiaramida}, {Chiavassa}, {Chornay}, {Comoretto}, {Contursi}, {Cooper}, {Cornez}, {Cowell}, {Crifo}, {Cropper}, {Crosta}, {Crowley}, {Dafonte}, {Dapergolas}, {David}, {David}, {de Laverny}, {De Luise}, \& {De March}}]{GaiaCollaboration2023A&A}
{Gaia Collaboration}, {Vallenari}, A., {Brown}, A.~G.~A., {et~al.} 2023{\natexlab{a}}, \aap, 674, A1, \dodoi{10.1051/0004-6361/202243940}

\bibitem[{{Gaia Collaboration} {et~al.}(2023{\natexlab{b}}){Gaia Collaboration}, {Arenou}, {Babusiaux}, {Barstow}, {Faigler}, {Jorissen}, {Kervella}, {Mazeh}, {Mowlavi}, {Panuzzo}, {Sahlmann}, {Shahaf}, {Sozzetti}, {Bauchet}, {Damerdji}, {Gavras}, {Giacobbe}, {Gosset}, {Halbwachs}, {Holl}, {Lattanzi}, {Leclerc}, {Morel}, {Pourbaix}, {Re Fiorentin}, {Sadowski}, {S{\'e}gransan}, {Siopis}, {Teyssier}, {Zwitter}, {Planquart}, {Brown}, {Vallenari}, {Prusti}, {de Bruijne}, {Biermann}, {Creevey}, {Ducourant}, {Evans}, {Eyer}, {Guerra}, {Hutton}, {Jordi}, {Klioner}, {Lammers}, {Lindegren}, {Luri}, {Mignard}, {Panem}, {Randich}, {Sartoretti}, {Soubiran}, {Tanga}, {Walton}, {Bailer-Jones}, {Bastian}, {Drimmel}, {Jansen}, {Katz}, {van Leeuwen}, {Bakker}, {Cacciari}, {Casta{\~n}eda}, {De Angeli}, {Fabricius}, {Fouesneau}, {Fr{\'e}mat}, {Galluccio}, {Guerrier}, {Heiter}, {Masana}, {Messineo}, {Nicolas}, {Nienartowicz}, {Pailler}, {Riclet}, {Roux}, {Seabroke}, {Sordo}, {Th{\'e}venin}, {Gracia-Abril}, {Portell}, {Altmann},
  {Andrae}, {Audard}, {Bellas-Velidis}, {Benson}, {Berthier}, {Blomme}, {Burgess}, {Busonero}, {Busso}, {C{\'a}novas}, {Carry}, {Cellino}, {Cheek}, {Clementini}, {Davidson}, {de Teodoro}, {Nu{\~n}ez Campos}, {Delchambre}, {Dell'Oro}, {Esquej}, {Fern{\'a}ndez-Hern{\'a}ndez}, {Fraile}, {Garabato}, {Garc{\'\i}a-Lario}, {Haigron}, {Hambly}, {Harrison}, {Hern{\'a}ndez}, {Hestroffer}, {Hodgkin}, {Jan{\ss}en}, {Jevardat de Fombelle}, {Jordan}, {Krone-Martins}, {Lanzafame}, {L{\"o}ffler}, {Marchal}, {Marrese}, {Moitinho}, {Muinonen}, {Osborne}, {Pancino}, {Pauwels}, {Recio-Blanco}, {Reyl{\'e}}, {Riello}, {Rimoldini}, {Roegiers}, {Rybizki}, {Sarro}, {Smith}, {Utrilla}, {van Leeuwen}, {Abbas}, {{\'A}brah{\'a}m}, {Abreu Aramburu}, {Aerts}, {Aguado}, {Ajaj}, {Aldea-Montero}, {Altavilla}, {{\'A}lvarez}, {Alves}, {Anders}, {Anderson}, {Anglada Varela}, {Antoja}, {Baines}, {Baker}, {Balaguer-N{\'u}{\~n}ez}, {Balbinot}, {Balog}, {Barache}, {Barbato}, {Barros}, {Bartolom{\'e}}, {Bassilana}, {Becciani}, {Bellazzini},
  {Berihuete}, {Bernet}, {Bertone}, {Bianchi}, {Binnenfeld}, {Blanco-Cuaresma}, {Blazere}, {Boch}, {Bombrun}, {Bossini}, {Bouquillon}, {Bragaglia}, {Bramante}, {Breedt}, {Bressan}, {Brouillet}, {Brugaletta}, {Bucciarelli}, {Burlacu}, {Butkevich}, {Buzzi}, {Caffau}, {Cancelliere}, {Cantat-Gaudin}, {Carballo}, {Carlucci}, {Carnerero}, {Carrasco}, {Casamiquela}, {Castellani}, {Castro-Ginard}, {Chaoul}, {Charlot}, {Chemin}, {Chiaramida}, {Chiavassa}, {Chornay}, \& {Comoretto}}]{GaiaCollaboration2023A&Ab}
{Gaia Collaboration}, {Arenou}, F., {Babusiaux}, C., {et~al.} 2023{\natexlab{b}}, \aap, 674, A34, \dodoi{10.1051/0004-6361/202243782}

\bibitem[{{Gao} \& {Li}(2023)}]{Gao2023MNRAS}
{Gao}, S.-J., \& {Li}, X.-D. 2023, \mnras, 525, 2605, \dodoi{10.1093/mnras/stad2446}

\bibitem[{{Garc{\'\i}a P{\'e}rez} {et~al.}(2016){Garc{\'\i}a P{\'e}rez}, {Allende Prieto}, {Holtzman}, {Shetrone}, {M{\'e}sz{\'a}ros}, {Bizyaev}, {Carrera}, {Cunha}, {Garc{\'\i}a-Hern{\'a}ndez}, {Johnson}, {Majewski}, {Nidever}, {Schiavon}, {Shane}, {Smith}, {Sobeck}, {Troup}, {Zamora}, {Weinberg}, {Bovy}, {Eisenstein}, {Feuillet}, {Frinchaboy}, {Hayden}, {Hearty}, {Nguyen}, {O'Connell}, {Pinsonneault}, {Wilson}, \& {Zasowski}}]{GarciaPerez2016AJ}
{Garc{\'\i}a P{\'e}rez}, A.~E., {Allende Prieto}, C., {Holtzman}, J.~A., {et~al.} 2016, \aj, 151, 144, \dodoi{10.3847/0004-6256/151/6/144}

\bibitem[{{Graham} {et~al.}(2019){Graham}, {Kulkarni}, {Bellm}, {Adams}, {Barbarino}, {Blagorodnova}, {Bodewits}, {Bolin}, {Brady}, {Cenko}, {Chang}, {Coughlin}, {De}, {Eadie}, {Farnham}, {Feindt}, {Franckowiak}, {Fremling}, {Gezari}, {Ghosh}, {Goldstein}, {Golkhou}, {Goobar}, {Ho}, {Huppenkothen}, {Ivezi{\'c}}, {Jones}, {Juric}, {Kaplan}, {Kasliwal}, {Kelley}, {Kupfer}, {Lee}, {Lin}, {Lunnan}, {Mahabal}, {Miller}, {Ngeow}, {Nugent}, {Ofek}, {Prince}, {Rauch}, {van Roestel}, {Schulze}, {Singer}, {Sollerman}, {Taddia}, {Yan}, {Ye}, {Yu}, {Barlow}, {Bauer}, {Beck}, {Belicki}, {Biswas}, {Brinnel}, {Brooke}, {Bue}, {Bulla}, {Burruss}, {Connolly}, {Cromer}, {Cunningham}, {Dekany}, {Delacroix}, {Desai}, {Duev}, {Feeney}, {Flynn}, {Frederick}, {Gal-Yam}, {Giomi}, {Groom}, {Hacopians}, {Hale}, {Helou}, {Henning}, {Hover}, {Hillenbrand}, {Howell}, {Hung}, {Imel}, {Ip}, {Jackson}, {Kaspi}, {Kaye}, {Kowalski}, {Kramer}, {Kuhn}, {Landry}, {Laher}, {Mao}, {Masci}, {Monkewitz}, {Murphy}, {Nordin}, {Patterson}, {Penprase},
  {Porter}, {Rebbapragada}, {Reiley}, {Riddle}, {Rigault}, {Rodriguez}, {Rusholme}, {van Santen}, {Shupe}, {Smith}, {Soumagnac}, {Stein}, {Surace}, {Szkody}, {Terek}, {Van Sistine}, {van Velzen}, {Vestrand}, {Walters}, {Ward}, {Zhang}, \& {Zolkower}}]{Graham19}
{Graham}, M.~J., {Kulkarni}, S.~R., {Bellm}, E.~C., {et~al.} 2019, \pasp, 131, 078001, \dodoi{10.1088/1538-3873/ab006c}

\bibitem[{{Green} {et~al.}(2019){Green}, {Schlafly}, {Zucker}, {Speagle}, \& {Finkbeiner}}]{Green2019ApJ}
{Green}, G.~M., {Schlafly}, E., {Zucker}, C., {Speagle}, J.~S., \& {Finkbeiner}, D. 2019, \apj, 887, 93, \dodoi{10.3847/1538-4357/ab5362}

\bibitem[{{Green} {et~al.}(1986){Green}, {Schmidt}, \& {Liebert}}]{Green1986ApJS}
{Green}, R.~F., {Schmidt}, M., \& {Liebert}, J. 1986, \apjs, 61, 305, \dodoi{10.1086/191115}

\bibitem[{{Gursky} {et~al.}(1971){Gursky}, {Kellogg}, {Murray}, {Leong}, {Tananbaum}, \& {Giacconi}}]{Gursky1971ApJL}
{Gursky}, H., {Kellogg}, E., {Murray}, S., {et~al.} 1971, \apjl, 167, L81, \dodoi{10.1086/180765}

\bibitem[{{Hallakoun} \& {Maoz}(2021)}]{Hallakoun2021MNRAS}
{Hallakoun}, N., \& {Maoz}, D. 2021, \mnras, 507, 398, \dodoi{10.1093/mnras/stab2145}

\bibitem[{{Hallakoun} {et~al.}(2024){Hallakoun}, {Shahaf}, {Mazeh}, {Toonen}, \& {Ben-Ami}}]{Hallakoun2024}
{Hallakoun}, N., {Shahaf}, S., {Mazeh}, T., {Toonen}, S., \& {Ben-Ami}, S. 2024, \apjl, 970, L11, \dodoi{10.3847/2041-8213/ad5e63}

\bibitem[{{Holberg} \& {Bergeron}(2006)}]{Holberg2006AJ}
{Holberg}, J.~B., \& {Bergeron}, P. 2006, \aj, 132, 1221, \dodoi{10.1086/505938}

\bibitem[{{Hui} {et~al.}(2018){Hui}, {Wu}, {Han}, {Kong}, \& {Tam}}]{Hui2018ApJ}
{Hui}, C.~Y., {Wu}, K., {Han}, Q., {Kong}, A.~K.~H., \& {Tam}, P.~H.~T. 2018, \apj, 864, 30, \dodoi{10.3847/1538-4357/aad5ec}

\bibitem[{{Hulse} \& {Taylor}(1975)}]{Hulse1975ApJL}
{Hulse}, R.~A., \& {Taylor}, J.~H. 1975, \apjl, 195, L51, \dodoi{10.1086/181708}

\bibitem[{{Iben} \& {Tutukov}(1984)}]{Iben1984ApJS}
{Iben}, Jr., I., \& {Tutukov}, A.~V. 1984, \apjs, 54, 335, \dodoi{10.1086/190932}

\bibitem[{{Kawaler}(1988)}]{Kawaler1988ApJ}
{Kawaler}, S.~D. 1988, \apj, 333, 236, \dodoi{10.1086/166740}

\bibitem[{{Koester}(2010)}]{Koester2010MmSAI}
{Koester}, D. 2010, \memsai, 81, 921

\bibitem[{{Lai} \& {Mu{\~n}oz}(2023)}]{Lai2023ARA&A}
{Lai}, D., \& {Mu{\~n}oz}, D.~J. 2023, \araa, 61, 517, \dodoi{10.1146/annurev-astro-052622-022933}

\bibitem[{{Liebert} {et~al.}(2005){Liebert}, {Bergeron}, \& {Holberg}}]{Liebert2005ApJS}
{Liebert}, J., {Bergeron}, P., \& {Holberg}, J.~B. 2005, \apjs, 156, 47, \dodoi{10.1086/425738}

\bibitem[{{Lin} {et~al.}(2026){Lin}, {Chen}, {Wang}, {Luo}, {Tang}, \& {Huang}}]{Lin2026ApJ}
{Lin}, J., {Chen}, H., {Wang}, B., {et~al.} 2026, \apj, 998, 341, \dodoi{10.3847/1538-4357/ae3c77}

\bibitem[{{Livio} \& {Soker}(1988)}]{Livio1988ApJ}
{Livio}, M., \& {Soker}, N. 1988, \apj, 329, 764, \dodoi{10.1086/166419}

\bibitem[{{Majewski} {et~al.}(2017){Majewski}, {Schiavon}, {Frinchaboy}, {Allende Prieto}, {Barkhouser}, {Bizyaev}, {Blank}, {Brunner}, {Burton}, {Carrera}, {Chojnowski}, {Cunha}, {Epstein}, {Fitzgerald}, {Garc{\'\i}a P{\'e}rez}, {Hearty}, {Henderson}, {Holtzman}, {Johnson}, {Lam}, {Lawler}, {Maseman}, {M{\'e}sz{\'a}ros}, {Nelson}, {Nguyen}, {Nidever}, {Pinsonneault}, {Shetrone}, {Smee}, {Smith}, {Stolberg}, {Skrutskie}, {Walker}, {Wilson}, {Zasowski}, {Anders}, {Basu}, {Beland}, {Blanton}, {Bovy}, {Brownstein}, {Carlberg}, {Chaplin}, {Chiappini}, {Eisenstein}, {Elsworth}, {Feuillet}, {Fleming}, {Galbraith-Frew}, {Garc{\'\i}a}, {Garc{\'\i}a-Hern{\'a}ndez}, {Gillespie}, {Girardi}, {Gunn}, {Hasselquist}, {Hayden}, {Hekker}, {Ivans}, {Kinemuchi}, {Klaene}, {Mahadevan}, {Mathur}, {Mosser}, {Muna}, {Munn}, {Nichol}, {O'Connell}, {Parejko}, {Robin}, {Rocha-Pinto}, {Schultheis}, {Serenelli}, {Shane}, {Silva Aguirre}, {Sobeck}, {Thompson}, {Troup}, {Weinberg}, \& {Zamora}}]{Majewski2017AJ}
{Majewski}, S.~R., {Schiavon}, R.~P., {Frinchaboy}, P.~M., {et~al.} 2017, \aj, 154, 94, \dodoi{10.3847/1538-3881/aa784d}

\bibitem[{{Marshall} {et~al.}(2006){Marshall}, {Robin}, {Reyl{\'e}}, {Schultheis}, \& {Picaud}}]{Marshall2006A&A}
{Marshall}, D.~J., {Robin}, A.~C., {Reyl{\'e}}, C., {Schultheis}, M., \& {Picaud}, S. 2006, \aap, 453, 635, \dodoi{10.1051/0004-6361:20053842}

\bibitem[{{Masci} {et~al.}(2019){Masci}, {Laher}, {Rusholme}, {Shupe}, {Groom}, {Surace}, {Jackson}, {Monkewitz}, {Beck}, {Flynn}, {Terek}, {Landry}, {Hacopians}, {Desai}, {Howell}, {Brooke}, {Imel}, {Wachter}, {Ye}, {Lin}, {Cenko}, {Cunningham}, {Rebbapragada}, {Bue}, {Miller}, {Mahabal}, {Bellm}, {Patterson}, {Juri{\'c}}, {Golkhou}, {Ofek}, {Walters}, {Graham}, {Kasliwal}, {Dekany}, {Kupfer}, {Burdge}, {Cannella}, {Barlow}, {Van Sistine}, {Giomi}, {Fremling}, {Blagorodnova}, {Levitan}, {Riddle}, {Smith}, {Helou}, {Prince}, \& {Kulkarni}}]{Masci19}
{Masci}, F.~J., {Laher}, R.~R., {Rusholme}, B., {et~al.} 2019, \pasp, 131, 018003, \dodoi{10.1088/1538-3873/aae8ac}

\bibitem[{{Mathieu} \& {Pols}(2025)}]{Mathieu2025ARA&A}
{Mathieu}, R.~D., \& {Pols}, O.~R. 2025, \araa, 63, 467, \dodoi{10.1146/annurev-astro-071221-054402}

\bibitem[{{Moe} \& {Di Stefano}(2017)}]{Moe2017ApJS}
{Moe}, M., \& {Di Stefano}, R. 2017, \apjs, 230, 15, \dodoi{10.3847/1538-4365/aa6fb6}

\bibitem[{{Moltzer} {et~al.}(2025){Moltzer}, {Pols}, {Van Winckel}, {Temmink}, \& {Wijdeveld}}]{Moltzer2025A&A}
{Moltzer}, C.~A.~S., {Pols}, O.~R., {Van Winckel}, H., {Temmink}, K.~D., \& {Wijdeveld}, M.~W. 2025, \aap, 703, A294, \dodoi{10.1051/0004-6361/202556437}

\bibitem[{{Morton}(2015)}]{Morton2015ascl}
{Morton}, T.~D. 2015, {isochrones: Stellar model grid package}, Astrophysics Source Code Library, record ascl:1503.010

\bibitem[{{Motherway} {et~al.}(2026){Motherway}, {Linck}, {Mathieu}, {Dixon}, {Stassun}, {Breivik}, {Majewski}, \& {Pols}}]{Motherway2026AJ}
{Motherway}, E., {Linck}, E., {Mathieu}, R.~D., {et~al.} 2026, \aj, 171, 159, \dodoi{10.3847/1538-3881/ae3b42}

\bibitem[{{Naoz}(2016)}]{Naoz2016ARA&A}
{Naoz}, S. 2016, \araa, 54, 441, \dodoi{10.1146/annurev-astro-081915-023315}

\bibitem[{{Naoz} {et~al.}(2013){Naoz}, {Farr}, {Lithwick}, {Rasio}, \& {Teyssandier}}]{Naoz2013MNRAS}
{Naoz}, S., {Farr}, W.~M., {Lithwick}, Y., {Rasio}, F.~A., \& {Teyssandier}, J. 2013, \mnras, 431, 2155, \dodoi{10.1093/mnras/stt302}

\bibitem[{{Nelemans} \& {Tout}(2005)}]{Nelemans2005MNRAS}
{Nelemans}, G., \& {Tout}, C.~A. 2005, \mnras, 356, 753, \dodoi{10.1111/j.1365-2966.2004.08496.x}

\bibitem[{{Nelemans} {et~al.}(2000){Nelemans}, {Verbunt}, {Yungelson}, \& {Portegies Zwart}}]{Nelemans2000A&A}
{Nelemans}, G., {Verbunt}, F., {Yungelson}, L.~R., \& {Portegies Zwart}, S.~F. 2000, \aap, 360, 1011, \dodoi{10.48550/arXiv.astro-ph/0006216}

\bibitem[{{Nelson} {et~al.}(2004){Nelson}, {Dubeau}, \& {MacCannell}}]{Nelson2004ApJ}
{Nelson}, L.~A., {Dubeau}, E., \& {MacCannell}, K.~A. 2004, \apj, 616, 1124, \dodoi{10.1086/421698}

\bibitem[{{Nidever} {et~al.}(2015){Nidever}, {Holtzman}, {Allende Prieto}, {Beland}, {Bender}, {Bizyaev}, {Burton}, {Desphande}, {Fleming}, {Garc{\'\i}a P{\'e}rez}, {Hearty}, {Majewski}, {M{\'e}sz{\'a}ros}, {Muna}, {Nguyen}, {Schiavon}, {Shetrone}, {Skrutskie}, {Sobeck}, \& {Wilson}}]{Nidever2015AJ}
{Nidever}, D.~L., {Holtzman}, J.~A., {Allende Prieto}, C., {et~al.} 2015, \aj, 150, 173, \dodoi{10.1088/0004-6256/150/6/173}

\bibitem[{{Nomoto}(1982)}]{Nomoto1982ApJ}
{Nomoto}, K. 1982, \apj, 253, 798, \dodoi{10.1086/159682}

\bibitem[{{Offner} {et~al.}(2023){Offner}, {Moe}, {Kratter}, {Sadavoy}, {Jensen}, \& {Tobin}}]{Offner2023ASPC}
{Offner}, S.~S.~R., {Moe}, M., {Kratter}, K.~M., {et~al.} 2023, in Astronomical Society of the Pacific Conference Series, Vol. 534, Protostars and Planets VII, ed. S.~{Inutsuka}, Y.~{Aikawa}, T.~{Muto}, K.~{Tomida}, \& M.~{Tamura}, 275, \dodoi{10.48550/arXiv.2203.10066}

\bibitem[{{Parkosidis} {et~al.}(2026{\natexlab{a}}){Parkosidis}, {Toonen}, {Dosopoulou}, \& {Laplace}}]{Parkosidis2026A&A_a}
{Parkosidis}, A., {Toonen}, S., {Dosopoulou}, F., \& {Laplace}, E. 2026{\natexlab{a}}, \aap, 706, A79, \dodoi{10.1051/0004-6361/202555096}

\bibitem[{{Parkosidis} {et~al.}(2026{\natexlab{b}}){Parkosidis}, {Toonen}, {Laplace}, \& {Dosopoulou}}]{Parkosidis2026A&A_b}
{Parkosidis}, A., {Toonen}, S., {Laplace}, E., \& {Dosopoulou}, F. 2026{\natexlab{b}}, \aap, 706, A357, \dodoi{10.1051/0004-6361/202558055}

\bibitem[{{Parkosidis} {et~al.}(2026{\natexlab{c}}){Parkosidis}, {Toonen}, {Laplace}, \& {Schaffenroth}}]{Parkosidis2026arXiv_c}
{Parkosidis}, A., {Toonen}, S., {Laplace}, E., \& {Schaffenroth}, V. 2026{\natexlab{c}}, arXiv e-prints, arXiv:2606.09464.
\newblock \doarXiv{2606.09464}

\bibitem[{{Peters} \& {Mathews}(1963)}]{Peters1963PhRv}
{Peters}, P.~C., \& {Mathews}, J. 1963, Physical Review, 131, 435, \dodoi{10.1103/PhysRev.131.435}

\bibitem[{{Phinney}(1992)}]{1992RSPTA.341...39P}
{Phinney}, E.~S. 1992, Philosophical Transactions of the Royal Society of London Series A, 341, 39, \dodoi{10.1098/rsta.1992.0084}

\bibitem[{{Radhakrishnan} \& {Srinivasan}(1982)}]{Radhakrishnan1982CSci}
{Radhakrishnan}, V., \& {Srinivasan}, G. 1982, Current Science, 51, 1096

\bibitem[{{Rappaport} {et~al.}(1982){Rappaport}, {Joss}, \& {Webbink}}]{Rappaport1982ApJ}
{Rappaport}, S., {Joss}, P.~C., \& {Webbink}, R.~F. 1982, \apj, 254, 616, \dodoi{10.1086/159772}

\bibitem[{{Rappaport} {et~al.}(1995){Rappaport}, {Podsiadlowski}, {Joss}, {Di Stefano}, \& {Han}}]{Rappaport1995MNRAS}
{Rappaport}, S., {Podsiadlowski}, P., {Joss}, P.~C., {Di Stefano}, R., \& {Han}, Z. 1995, \mnras, 273, 731, \dodoi{10.1093/mnras/273.3.731}

\bibitem[{{Rappaport} {et~al.}(1983){Rappaport}, {Verbunt}, \& {Joss}}]{Rappaport1983ApJ}
{Rappaport}, S., {Verbunt}, F., \& {Joss}, P.~C. 1983, \apj, 275, 713, \dodoi{10.1086/161569}

\bibitem[{{Rebassa-Mansergas} {et~al.}(2010){Rebassa-Mansergas}, {G{\"a}nsicke}, {Schreiber}, {Koester}, \& {Rodr{\'\i}guez-Gil}}]{Rebassa-Mansergas2010MNRAS}
{Rebassa-Mansergas}, A., {G{\"a}nsicke}, B.~T., {Schreiber}, M.~R., {Koester}, D., \& {Rodr{\'\i}guez-Gil}, P. 2010, \mnras, 402, 620, \dodoi{10.1111/j.1365-2966.2009.15915.x}

\bibitem[{{Rebassa-Mansergas} {et~al.}(2012){Rebassa-Mansergas}, {Nebot G{\'o}mez-Mor{\'a}n}, {Schreiber}, {G{\"a}nsicke}, {Schwope}, {Gallardo}, \& {Koester}}]{Rebassa-Mansergas2012MNRAS}
{Rebassa-Mansergas}, A., {Nebot G{\'o}mez-Mor{\'a}n}, A., {Schreiber}, M.~R., {et~al.} 2012, \mnras, 419, 806, \dodoi{10.1111/j.1365-2966.2011.19923.x}

\bibitem[{{Reiners} {et~al.}(2009){Reiners}, {Basri}, \& {Browning}}]{Reiners2009ApJ}
{Reiners}, A., {Basri}, G., \& {Browning}, M. 2009, \apj, 692, 538, \dodoi{10.1088/0004-637X/692/1/538}

\bibitem[{{Riley} {et~al.}(2022){Riley}, {Agrawal}, {Barrett}, {Boyett}, {Broekgaarden}, {Chattopadhyay}, {Gaebel}, {Gittins}, {Hirai}, {Howitt}, {Justham}, {Khandelwal}, {Kummer}, {Lau}, {Mandel}, {de Mink}, {Neijssel}, {Riley}, {van Son}, {Stevenson}, {Vigna-G{\'o}mez}, {Vinciguerra}, {Wagg}, {Willcox}, \& {Team Compas}}]{Riley2022ApJS}
{Riley}, J., {Agrawal}, P., {Barrett}, J.~W., {et~al.} 2022, \apjs, 258, 34, \dodoi{10.3847/1538-4365/ac416c}

\bibitem[{{Romero} {et~al.}(2012){Romero}, {C{\'o}rsico}, {Althaus}, {Kepler}, {Castanheira}, \& {Miller Bertolami}}]{Romero2012MNRAS}
{Romero}, A.~D., {C{\'o}rsico}, A.~H., {Althaus}, L.~G., {et~al.} 2012, \mnras, 420, 1462, \dodoi{10.1111/j.1365-2966.2011.20134.x}

\bibitem[{{Romero} {et~al.}(2019){Romero}, {Kepler}, {Joyce}, {Lauffer}, \& {C{\'o}rsico}}]{Romero2019MNRAS}
{Romero}, A.~D., {Kepler}, S.~O., {Joyce}, S.~R.~G., {Lauffer}, G.~R., \& {C{\'o}rsico}, A.~H. 2019, \mnras, 484, 2711, \dodoi{10.1093/mnras/stz160}

\bibitem[{{Schatzman}(1962)}]{Schatzman1962AnAp}
{Schatzman}, E. 1962, Annales d'Astrophysique, 25, 18

\bibitem[{{Scherbak} \& {Fuller}(2023)}]{Scherbak2023MNRAS}
{Scherbak}, P., \& {Fuller}, J. 2023, \mnras, 518, 3966, \dodoi{10.1093/mnras/stac3313}

\bibitem[{{Schreiber} \& {G{\"a}nsicke}(2003)}]{Schreiber2003A&A}
{Schreiber}, M.~R., \& {G{\"a}nsicke}, B.~T. 2003, \aap, 406, 305, \dodoi{10.1051/0004-6361:20030801}

\bibitem[{{Schreiber} {et~al.}(2010){Schreiber}, {G{\"a}nsicke}, {Rebassa-Mansergas}, {Nebot Gomez-Moran}, {Southworth}, {Schwope}, {M{\"u}ller}, {Papadaki}, {Pyrzas}, {Rabitz}, {Rodr{\'\i}guez-Gil}, {Schmidtobreick}, {Schwarz}, {Tappert}, {Toloza}, {Vogel}, \& {Zorotovic}}]{Schreiber2010A&A}
{Schreiber}, M.~R., {G{\"a}nsicke}, B.~T., {Rebassa-Mansergas}, A., {et~al.} 2010, \aap, 513, L7, \dodoi{10.1051/0004-6361/201013990}

\bibitem[{{Shahaf} {et~al.}(2024){Shahaf}, {Hallakoun}, {Mazeh}, {Ben-Ami}, {Rekhi}, {El-Badry}, \& {Toonen}}]{Shahaf2024MNRAS}
{Shahaf}, S., {Hallakoun}, N., {Mazeh}, T., {et~al.} 2024, \mnras, 529, 3729, \dodoi{10.1093/mnras/stae773}

\bibitem[{{Shahaf} {et~al.}(2019){Shahaf}, {Mazeh}, {Faigler}, \& {Holl}}]{Shahaf2019MNRAS}
{Shahaf}, S., {Mazeh}, T., {Faigler}, S., \& {Holl}, B. 2019, \mnras, 487, 5610, \dodoi{10.1093/mnras/stz1636}

\bibitem[{{Shariat} \& {El-Badry}(2026{\natexlab{a}})}]{Shariat2026PASP}
{Shariat}, C., \& {El-Badry}, K. 2026{\natexlab{a}}, \pasp, 138, 034202, \dodoi{10.1088/1538-3873/ae453b}

\bibitem[{{Shariat} \& {El-Badry}(2026{\natexlab{b}})}]{Shariat2026arXiv}
---. 2026{\natexlab{b}}, \pasp, 138, 034202, \dodoi{10.1088/1538-3873/ae453b}

\bibitem[{{Shariat} {et~al.}(2023){Shariat}, {Naoz}, {Hansen}, {Angelo}, {Michaely}, \& {Stephan}}]{Shariat2023ApJL}
{Shariat}, C., {Naoz}, S., {Hansen}, B. M.~S., {et~al.} 2023, \apjl, 955, L14, \dodoi{10.3847/2041-8213/acf76b}

\bibitem[{{Sharma} {et~al.}(2011){Sharma}, {Bland-Hawthorn}, {Johnston}, \& {Binney}}]{Sharma2011ApJ}
{Sharma}, S., {Bland-Hawthorn}, J., {Johnston}, K.~V., \& {Binney}, J. 2011, \apj, 730, 3, \dodoi{10.1088/0004-637X/730/1/3}

\bibitem[{{Shetrone} {et~al.}(2015){Shetrone}, {Bizyaev}, {Lawler}, {Allende Prieto}, {Johnson}, {Smith}, {Cunha}, {Holtzman}, {Garc{\'\i}a P{\'e}rez}, {M{\'e}sz{\'a}ros}, {Sobeck}, {Zamora}, {Garc{\'\i}a-Hern{\'a}ndez}, {Souto}, {Chojnowski}, {Koesterke}, {Majewski}, \& {Zasowski}}]{Shetrone2015ApJS}
{Shetrone}, M., {Bizyaev}, D., {Lawler}, J.~E., {et~al.} 2015, \apjs, 221, 24, \dodoi{10.1088/0067-0049/221/2/24}

\bibitem[{{Shi} {et~al.}(2026){Shi}, {Ge}, {Li}, {Belloni}, {Zheng}, {Jiang}, {Chen}, {Santos-Garcia}, {Torres}, {Rebassa-Mansergas}, {Chen}, \& {Han}}]{Shi2026arXiv}
{Shi}, Y., {Ge}, H., {Li}, Z., {et~al.} 2026, \apj, 1007, 15, \dodoi{10.3847/1538-4357/ae876a}

\bibitem[{{Sills} {et~al.}(2000){Sills}, {Pinsonneault}, \& {Terndrup}}]{Sills2000ApJ}
{Sills}, A., {Pinsonneault}, M.~H., \& {Terndrup}, D.~M. 2000, \apj, 534, 335, \dodoi{10.1086/308739}

\bibitem[{{Skumanich}(1972)}]{Skumanich1972ApJ}
{Skumanich}, A. 1972, \apj, 171, 565, \dodoi{10.1086/151310}

\bibitem[{{Smith} {et~al.}(2021){Smith}, {Bizyaev}, {Cunha}, {Shetrone}, {Souto}, {Allende Prieto}, {Masseron}, {M{\'e}sz{\'a}ros}, {J{\"o}nsson}, {Hasselquist}, {Osorio}, {Garc{\'\i}a-Hern{\'a}ndez}, {Plez}, {Beaton}, {Holtzman}, {Majewski}, {Stringfellow}, \& {Sobeck}}]{Smith2021AJ}
{Smith}, V.~V., {Bizyaev}, D., {Cunha}, K., {et~al.} 2021, \aj, 161, 254, \dodoi{10.3847/1538-3881/abefdc}

\bibitem[{{Soberman} {et~al.}(1997){Soberman}, {Phinney}, \& {van den Heuvel}}]{Soberman1997A&A}
{Soberman}, G.~E., {Phinney}, E.~S., \& {van den Heuvel}, E.~P.~J. 1997, \aap, 327, 620, \dodoi{10.48550/arXiv.astro-ph/9703016}

\bibitem[{{Tang} {et~al.}(2014){Tang}, {Bressan}, {Rosenfield}, {Slemer}, {Marigo}, {Girardi}, \& {Bianchi}}]{Tang2014MNRAS}
{Tang}, J., {Bressan}, A., {Rosenfield}, P., {et~al.} 2014, \mnras, 445, 4287, \dodoi{10.1093/mnras/stu2029}

\bibitem[{{Tauris} \& {Savonije}(1999)}]{Tauris1999A&A}
{Tauris}, T.~M., \& {Savonije}, G.~J. 1999, \aap, 350, 928, \dodoi{10.48550/arXiv.astro-ph/9909147}

\bibitem[{{Temmink} {et~al.}(2023){Temmink}, {Pols}, {Justham}, {Istrate}, \& {Toonen}}]{Temmink2023AA}
{Temmink}, K.~D., {Pols}, O.~R., {Justham}, S., {Istrate}, A.~G., \& {Toonen}, S. 2023, \aap, 669, A45, \dodoi{10.1051/0004-6361/202244137}

\bibitem[{{Toonen} {et~al.}(2016){Toonen}, {Hamers}, \& {Portegies Zwart}}]{Toonen2016ComAC}
{Toonen}, S., {Hamers}, A., \& {Portegies Zwart}, S. 2016, Computational Astrophysics and Cosmology, 3, 6, \dodoi{10.1186/s40668-016-0019-0}

\bibitem[{{Toonen} {et~al.}(2020){Toonen}, {Portegies Zwart}, {Hamers}, \& {Bandopadhyay}}]{Toonen2020A&A}
{Toonen}, S., {Portegies Zwart}, S., {Hamers}, A.~S., \& {Bandopadhyay}, D. 2020, \aap, 640, A16, \dodoi{10.1051/0004-6361/201936835}

\bibitem[{{Torres} {et~al.}(2025){Torres}, {Gili}, {Rebassa-Mansergas}, {Santos-Garc{\'\i}a}, {Brown}, \& {Parsons}}]{Torres2025A&A}
{Torres}, S., {Gili}, M., {Rebassa-Mansergas}, A., {et~al.} 2025, \aap, 698, A173, \dodoi{10.1051/0004-6361/202554039}

\bibitem[{{van Son} {et~al.}(2026){van Son}, {Yamaguchi}, {Nagarajan}, {Shenar}, {Sen}, {Laroche}, {Leiner}, {Sana}, \& {Pols}}]{vanSon2026arXiv}
{van Son}, L. A.~C., {Yamaguchi}, N., {Nagarajan}, P., {et~al.} 2026, \apjs, 286, 13, \dodoi{10.3847/1538-4365/ae8d04}

\bibitem[{{Verbunt} \& {Zwaan}(1981)}]{Verbunt1981A&A}
{Verbunt}, F., \& {Zwaan}, C. 1981, \aap, 100, L7

\bibitem[{{Webbink}(1984)}]{Webbink1984ApJ}
{Webbink}, R.~F. 1984, \apj, 277, 355, \dodoi{10.1086/161701}

\bibitem[{{Weber} \& {Davis}(1967)}]{Weber1967ApJ}
{Weber}, E.~J., \& {Davis}, Jr., L. 1967, \apj, 148, 217, \dodoi{10.1086/149138}

\bibitem[{{Weidemann}(2000)}]{Weidemann2000AA}
{Weidemann}, V. 2000, \aap, 363, 647

\bibitem[{{Whelan} \& {Iben}(1973)}]{Whelan1973ApJ}
{Whelan}, J., \& {Iben}, Jr., I. 1973, \apj, 186, 1007, \dodoi{10.1086/152565}

\bibitem[{{Willems} \& {Kolb}(2004)}]{Willems2004A&A}
{Willems}, B., \& {Kolb}, U. 2004, \aap, 419, 1057, \dodoi{10.1051/0004-6361:20040085}

\bibitem[{{Wilson} {et~al.}(2019){Wilson}, {Hearty}, {Skrutskie}, {Majewski}, {Holtzman}, {Eisenstein}, {Gunn}, {Blank}, {Henderson}, {Smee}, {Nelson}, {Nidever}, {Arns}, {Barkhouser}, {Barr}, {Beland}, {Bershady}, {Blanton}, {Brunner}, {Burton}, {Carey}, {Carr}, {Colque}, {Crane}, {Damke}, {Davidson}, {Dean}, {Di Mille}, {Don}, {Ebelke}, {Evans}, {Fitzgerald}, {Gillespie}, {Hall}, {Harding}, {Harding}, {Hammond}, {Hancock}, {Harrison}, {Hope}, {Horne}, {Karakla}, {Lam}, {Leger}, {MacDonald}, {Maseman}, {Matsunari}, {Melton}, {Mitcheltree}, {O'Brien}, {O'Connell}, {Patten}, {Richardson}, {Rieke}, {Rieke}, {Roman-Lopes}, {Schiavon}, {Sobeck}, {Stolberg}, {Stoll}, {Tembe}, {Trujillo}, {Uomoto}, {Vernieri}, {Walker}, {Weinberg}, {Young}, {Anthony-Brumfield}, {Bizyaev}, {Breslauer}, {De Lee}, {Downey}, {Halverson}, {Huehnerhoff}, {Klaene}, {Leon}, {Long}, {Mahadevan}, {Malanushenko}, {Nguyen}, {Owen}, {S{\'a}nchez-Gallego}, {Sayres}, {Shane}, {Shectman}, {Shetrone}, {Skinner}, {Stauffer}, \&
  {Zhao}}]{Wilson2019PASP}
{Wilson}, J.~C., {Hearty}, F.~R., {Skrutskie}, M.~F., {et~al.} 2019, \pasp, 131, 055001, \dodoi{10.1088/1538-3873/ab0075}

\bibitem[{{Wonnacott} {et~al.}(1993){Wonnacott}, {Kellett}, \& {Stickland}}]{Wonnacott1993MNRAS}
{Wonnacott}, D., {Kellett}, B.~J., \& {Stickland}, D.~J. 1993, \mnras, 262, 277, \dodoi{10.1093/mnras/262.2.277}

\bibitem[{{Wu} {et~al.}(2025){Wu}, {Hadden}, {Dewberry}, {El-Badry}, \& {Matzner}}]{Wu2025ApJL}
{Wu}, Y., {Hadden}, S., {Dewberry}, J., {El-Badry}, K., \& {Matzner}, C.~D. 2025, \apjl, 982, L34, \dodoi{10.3847/2041-8213/adb751}

\bibitem[{{Yamaguchi} {et~al.}(2024{\natexlab{a}}){Yamaguchi}, {El-Badry}, {Rees}, {Shahaf}, {Mazeh}, \& {Andrae}}]{Yamaguchi2024PASP}
{Yamaguchi}, N., {El-Badry}, K., {Rees}, N.~R., {et~al.} 2024{\natexlab{a}}, \pasp, 136, 084202, \dodoi{10.1088/1538-3873/ad6809}

\bibitem[{{Yamaguchi} {et~al.}(2025){Yamaguchi}, {El-Badry}, \& {Shahaf}}]{Yamaguchi2025PASP}
{Yamaguchi}, N., {El-Badry}, K., \& {Shahaf}, S. 2025, \pasp, 137, 104205, \dodoi{10.1088/1538-3873/ae0d30}

\bibitem[{{Yamaguchi} {et~al.}(2024{\natexlab{b}}){Yamaguchi}, {El-Badry}, {Fuller}, {Latham}, {Cargile}, {Mazeh}, {Shahaf}, {Bieryla}, {Buchhave}, \& {Hobson}}]{Yamaguchi2024MNRAS}
{Yamaguchi}, N., {El-Badry}, K., {Fuller}, J., {et~al.} 2024{\natexlab{b}}, \mnras, 527, 11719, \dodoi{10.1093/mnras/stad4005}

\bibitem[{{Yuan} {et~al.}(2013){Yuan}, {Liu}, \& {Xiang}}]{Yuan2013MNRAS}
{Yuan}, H.~B., {Liu}, X.~W., \& {Xiang}, M.~S. 2013, \mnras, 430, 2188, \dodoi{10.1093/mnras/stt039}

\bibitem[{{Zahn}(1977)}]{Zahn1977A&A}
{Zahn}, J.-P. 1977, \aap, 57, 383

\bibitem[{{Zhang} {et~al.}(2023){Zhang}, {Green}, \& {Rix}}]{Zhang2023MNRAS}
{Zhang}, X., {Green}, G.~M., \& {Rix}, H.-W. 2023, \mnras, 524, 1855, \dodoi{10.1093/mnras/stad1941}

\bibitem[{{Zorotovic} {et~al.}(2010){Zorotovic}, {Schreiber}, {G{\"a}nsicke}, \& {Nebot G{\'o}mez-Mor{\'a}n}}]{Zorotovic2010A&A}
{Zorotovic}, M., {Schreiber}, M.~R., {G{\"a}nsicke}, B.~T., \& {Nebot G{\'o}mez-Mor{\'a}n}, A. 2010, \aap, 520, A86, \dodoi{10.1051/0004-6361/200913658}

\end{thebibliography}



\end{document}